\documentclass[aps,pre,amsmath,amsfonts,amssymb,floatfix,superscriptaddress,nofootinbib]{revtex4}
\usepackage{mathrsfs} \usepackage{graphicx,color} \usepackage{bbm} \usepackage{multirow}

\let\a=\alpha \let\b=\beta \let\g=\gamma \let\d=\delta
   
\let\l=\lambda    
  \let\f=\varphi 
 \let\y=\upsilon  \let\G=\Gamma
\let\D=\Delta \let\Th=\Theta\let\L=\Lambda  
   
 \let\r=\rho  \let\io=\infty

\def\FF{{\cal F}} 
 
\def\LL{{\cal L}}  
\def\DD{{\cal D}}\def\GG{{\cal G}} \def\SS{{\cal S}}

  \def\erf{\text{erf}}

\def\de{\mathrm d}

\def\to{\rightarrow} \def\la{\left\langle} \def\ra{\right\rangle}

\newcommand{\beq}{\begin{equation}} \newcommand{\eeq}{\end{equation}}
\newcommand{\wh}{\widehat} 
\newcommand{\Tr}{\text{Tr}}

\begin{document}

\title{
Exact theory of dense amorphous hard spheres in high dimension \\
II. The high density regime and the Gardner transition
} 

\author{Jorge Kurchan}
\affiliation{
LPS,
\'Ecole Normale Sup\'erieure, UMR 8550 CNRS, 24 Rue Lhomond, 75005 France
}

\author{Giorgio Parisi}
\affiliation{Dipartimento di Fisica,
Sapienza Universit\'a di Roma,
INFN, Sezione di Roma I, IPFC -- CNR,
P.le A. Moro 2, I-00185 Roma, Italy
}

\author{Pierfrancesco Urbani}
\affiliation{Dipartimento di Fisica,
Sapienza Universit\'a di Roma,
INFN, Sezione di Roma I, IPFC -- CNR,
P.le A. Moro 2, I-00185 Roma, Italy
}
\affiliation{Laboratoire de Physique Th\'eorique et Mod\`eles
    Statistiques, CNRS et Universit\'e Paris-Sud 11,
UMR8626, B\^at. 100, 91405 Orsay Cedex, France}

\author{Francesco Zamponi}
\affiliation{LPT,
\'Ecole Normale Sup\'erieure, UMR 8549 CNRS, 24 Rue Lhomond, 75005 France}

\begin{abstract}
\centerline{\bf Abstract}
\smallskip
We consider the theory of the glass phase and jamming of hard spheres in the large space dimension limit.
Building upon the exact expression for the free-energy functional obtained previously, we find that the 
Random First Order Transition (RFOT) scenario is realized here with two thermodynamic transitions: 
the usual Kauzmann point associated with entropy crisis,
and a further transition at higher pressures in which  a glassy structure of micro-states is developed 
within each amorphous state.
This kind of glass-glass  transition into a phase dominating the higher densities was described years 
ago by Elisabeth Gardner, 
and may well be a generic feature of RFOT. 
Micro states that are small excitations of  an amorphous matrix -- separated by low entropic or energetic barriers -- 
thus emerge naturally, 
and modify the high pressure (or low temperature) limit of the thermodynamic functions. 
\end{abstract}

\maketitle

Keywords: glass transition; jamming; Gardner transition; spin glasses

\section{Introduction}

The Random First Order Transition (RFOT) scenario for glasses introduced by 
Kirkpatrick, Thirumalai and Wolynes~\cite{KW87b,KT87,KTW89,WL12} 
proposes that the glass transition is represented 
-- at least at the mean-field level -- by a freezing transition similar to the one of the Random Energy Model~\cite{De81}, 
described mathematically by a one-step replica symmetry breaking (1RSB) ansatz~\cite{MPV87}.
Although this scenario was first based on an analogy with spin glasses~\cite{KW87b,KT87}, it was quickly
realized in a pioneering work by Kirkpatrick and Wolynes~\cite{KW87} 
that the glass transition of $d$-dimensional hard spheres 
in the limit $d\to\io$ could be described within the same framework. Much later~\cite{PZ06a,PZ10}, 
similar results were obtained
using the replica method, which also allowed for a detailed description of the glass phase and in particular
of the {\it jamming} point where the pressure of the glass becomes infinite, corresponding to its close packing --
which was therefore called {\it glass close packing} (GCP) and is closely related to the random close packing
concept introduced by Bernal much earlier~\cite{BM60}.

The main advantage of the replica method is that it allows for a unified treatment of both the glass and 
the jamming transitions, within a simple static RFOT scenario, and it also allows for a partial understanding of 
dynamical aspects. Moreover, there is hope that the replica method can provide an {\it exact} result in the 
limit $d\to\io$. A first step in this direction was performed in the first paper of this series~\cite{KPZ12},
where we have shown that the thermodynamics of hard spheres in the
limit of high dimensions may be exactly obtained from the knowledge of the distribution of the two-point 
correlation function between states, encoded in the Parisi parameter. In the same paper it was shown that 
once a 1RSB ansatz is made, one recovers exactly the Gaussian replica free energy that was used in Ref.~\cite{PZ06a,PZ10}
to derive estimates of the various transitions that characterize the RFOT scenario at the 1RSB level.

In this paper we take a second important step, by investigating the stability of the 1RSB solution towards further
levels of replica symmetry breaking.
We find that for higher pressures, well above the RFOT one, there is 
a second transition (a so-called Gardner transition~\cite{Ga85}) leading to a somewhat different physics in that limit, 
and in particular around the jamming point. We believe that this physics is intimately connected with the peculiar
mechanical properties of jammed states of hard spheres, that have been recently characterized in much 
detail~\cite{LNSW10,He10}.

The rest of the paper is organized as follows. We start our presentation by a general discussion of the RFOT
scenario, of its connection with the physics of jamming, and of the main new features that are due to the
presence of the Gardner transition. This discussion is reported in Sec.~\ref{sec:general} and it is for the moment mostly
speculative, although some parts of it have been previously studied in spin glass models. 
Next, we present our new results, which constitute a first important step to substantiate this picture for hard
spheres in the $d\to\io$ limit. 
In Sec.~\ref{sec:II} we provide a proof of the correctness of the Gaussian ansatz for a generic form of the
overlap matrix, extending the main result of Ref.~\cite{KPZ12};
in Sec.~\ref{sec:IIa} we recall a few important results of Ref.~\cite{PZ10,KPZ12} 
that are directly needed here; 
in Sec.~\ref{sec:III} we present our main results for the Hessian matrix 
of the 1RSB solution and in particular its so-called {\it replicon} eigenvalue, 
that is responsible for the instability of this solution (i.e. the Gardner transition);
in Sec.~\ref{app:MCT} we discuss the cubic terms in the expansion around the 1RSB solution and from them
we extract the dynamical exponents that characterize the glass transition;
in Sec.~\ref{sec:IV} we present an approximate calculation to obtain an order of magnitude for the
Gardner transition pressure in finite $d$; in Sec.~\ref{sec:V} we summarize and draw our conclusions.

\section{A general RFOT scenario for the glass and jamming transitions}
\label{sec:general}

 \begin{figure}
\includegraphics[width=.5\textwidth]{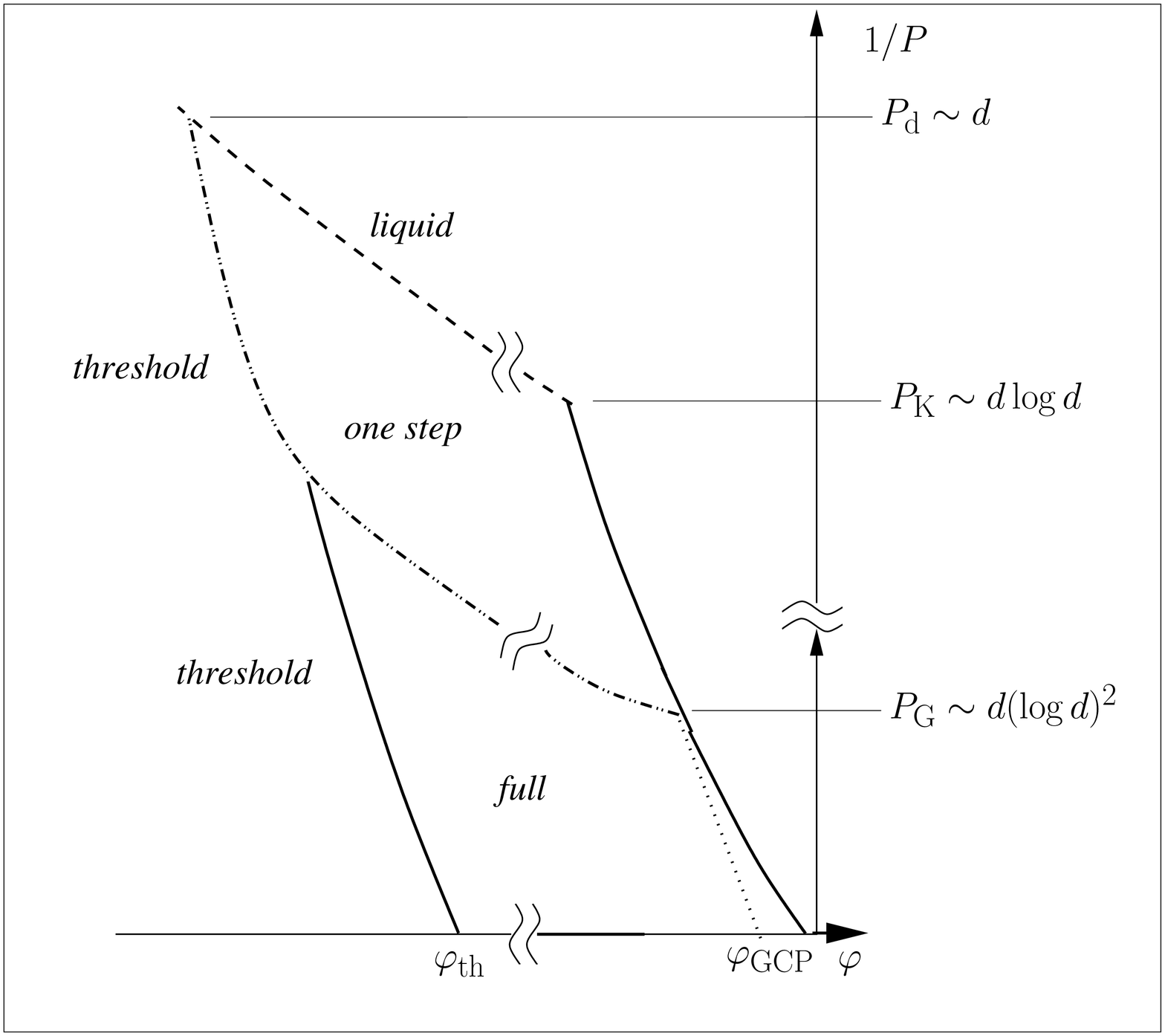}
\caption{
A sketch of the phase diagram.
}
\label{pspi}
\end{figure}

\subsection{The generic phase diagram of RFOT models}

\begin{figure}[t]
\includegraphics[width=.5\textwidth]{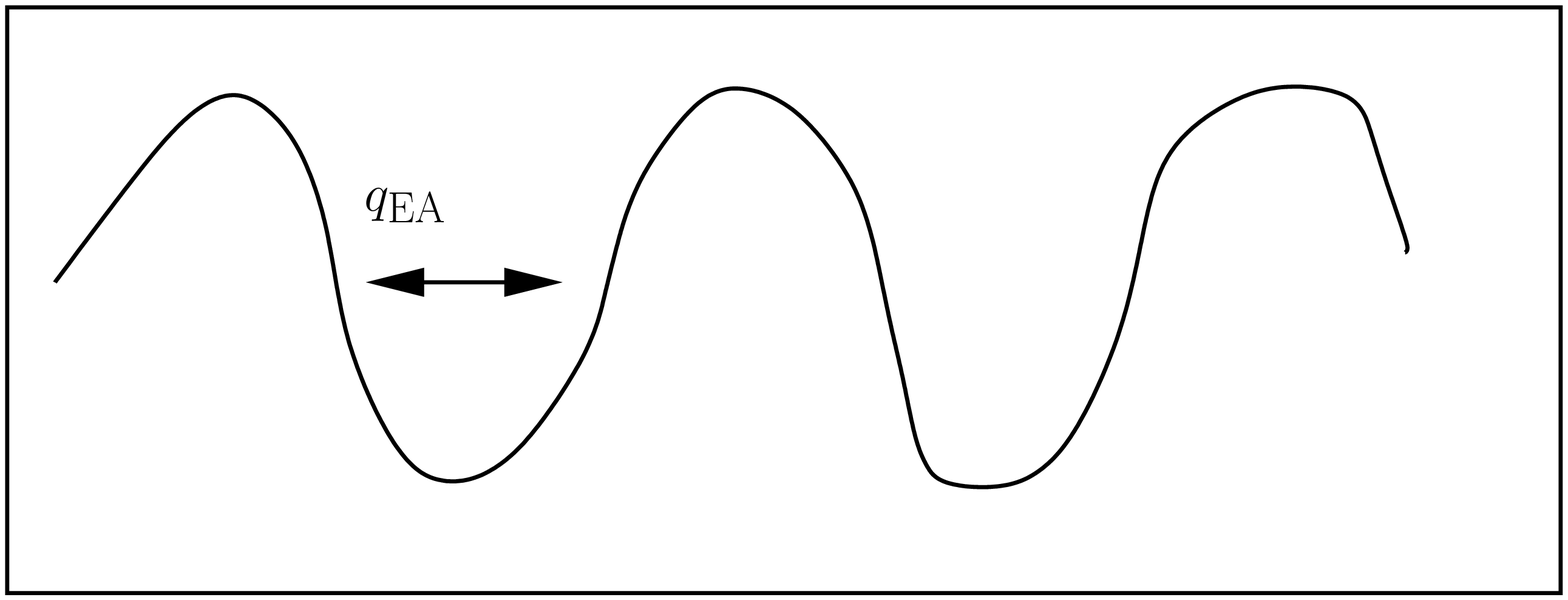}
\includegraphics[width=.5\textwidth]{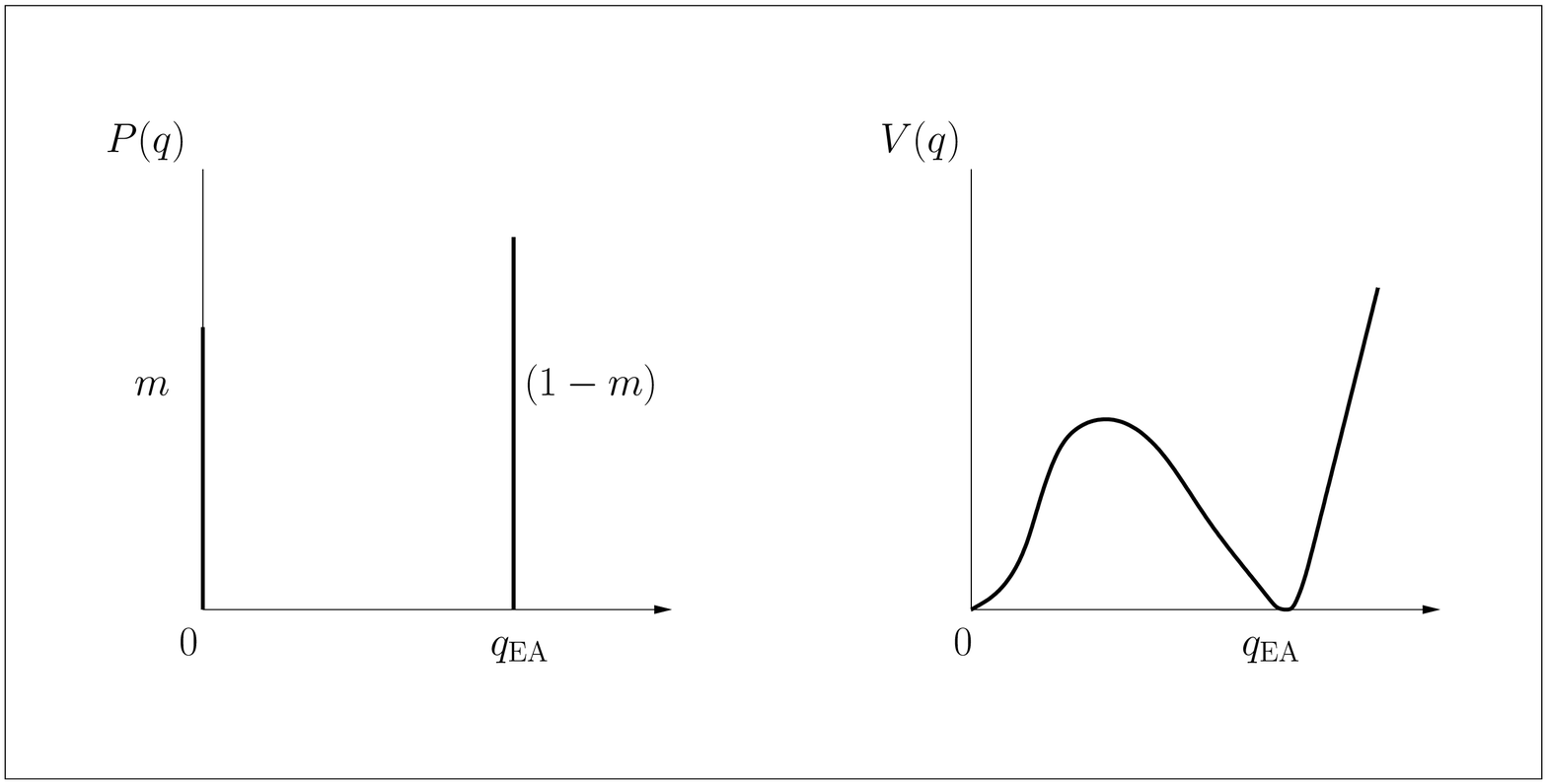}
\caption{
A sketch for equilibrium $T_{\rm G} \le T \le T_k$ or $P_{\rm G} \ge P \ge P_{\rm K}$.
Top: a cartoon of the free energy and its minima of width $q_{\rm EA}$.
Bottom: the Parisi order parameter $P(q)$ and the effective potential $V(q)$.
}
\label{pq1}
 \label{well1}
\end{figure}

As is by now well known, Kirkpatrick, Thirumalai and Wolynes' scenario for the   
liquid-glass transition involves a first point at which the equilibrium state fractures into an exponential number of
ergodic components: this is the dynamical temperature $T_{\rm d}$ (or pressure $P_{\rm d}$, see Fig.~\ref{pspi}),
also called Mode-Coupling temperature
because in low dimensions it can be computed using Mode-Coupling theory.  
The ergodic components are only truly dynamically separated in the mean-field limit, while in a 
realistic short-range finite-dimensional situation the system is still ergodic, although the dynamics slows down.
As the temperature is lowered, or the pressure increased, the number of metastable states 
contributing to equilibrium diminishes,
until a point is reached where the equilibrium system is left with only the deepest amorphous states: 
this is the Kauzmann point,
beyond which the thermodynamics stays dominated by (or ``frozen in'') those states.
From a purely equilibrium point of view, one may picture the situation at  $P>P_{\rm K}$ (or $T<T_{\rm K}$) as in the 
sketch of Fig. \ref{well1}, with widely separated states of ``size'' $q$, defined for example
as:
\begin{equation}
q = \frac{1}{N} \sum_{i} \cos[ k \cdot (r_i^a- r_i^b)]
\end{equation}
with $a,b$ two copies (replicas) of the system and
$k$ a vector of length comparable to the inverse of the inter-particle distance
(alternative definitions of $q$ are possible, see~\cite{FJPUZ13} for a review).
A more precise way of stating the same thing is to introduce the effective potential $V(q)$~\cite{FP}, 
which counts the logarithm of the 
number of configurations having correlation exactly $q$ with a ``reference'' equilibrium configuration. One obtains a picture
as in Fig. \ref{pq1}, where one sees that configurations are either close, with overlap $q=q_{\rm EA}$ 
(with probability $1-m$), or very far away -- in other states, with overlap $q=0$ --
with probability $m$, corresponding to the two minima in the effective potential 
and to the two peaks in the Parisi order parameter $P(q)$. 
Note that $P(q)$ may in general only be non-zero where $V(q)$ takes the minimal value, and that
$q_{\rm EA}$ plays the role of the Edwards-Anderson order parameter~\cite{MPV87}.

\begin{figure}[t]
\includegraphics[width=.5\textwidth]{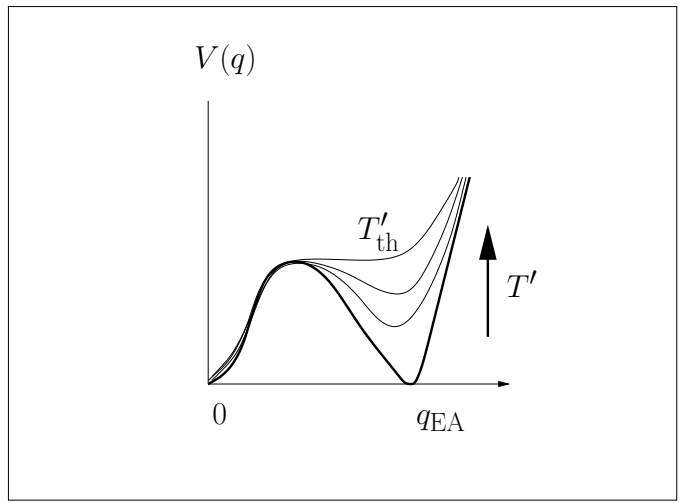}
\caption{Effective potential associated to higher  metastable states. Metastability disappears at the threshold level.}
\label{pq4}
\end{figure}

This construction concerns equilibrium configurations, but may be generalized to describe metastable states~\cite{KT89,Mo95} by
choosing the ``reference'' configuration, rather than from equilibrium, from a system perturbed by a small ``pinning field'',  itself  
thermalized at a higher temperature $T'=T/m$. Technically speaking, following Monasson~\cite{Mo95}, this amounts to the following
calculation: one considers $m$ weakly coupled replicas at temperature $T$, takes an equilibrium configurations of one of the replicas
as the reference configuration, and couples to it an additional replica who is forced to stay at distance $q$ from it. Then one computes
the free energy of the additional replica, and averages it over the other $m$. This amounts to fixing the Parisi parameter $m$, rather
than choosing the value that maximizes the free energy.
In this way, one obtains a bistable form for the effective potential, up to a threshold value $T'_{th}=T/m_{th}$ at
which the minimum close to $q_{\rm EA}$ disappears (Fig.~\ref{pq4}), and at precisely the threshold
level, the stability matrix corresponding to the minimum at $q_{\rm EA}$ develops zero modes, signaling the fact that the 
states close to the threshold level are {\em marginal}. 
This shows up within the replica scenario as the vanishing of the ``replicon'' eigenvalue \cite{DeKo}, 
and within the Thouless-Anderson-Palmer approach~\cite{MPV87} as the free-energy Hessian developing zero eigenvalues.
A crucial result of Ref.~\cite{CK93} is that 
the out of equilibrium aging dynamics happens exactly at this threshold level, and it exploits the marginality of the 
threshold states to explore phase space. 

\begin{figure}[t]
\includegraphics[width=.5\textwidth]{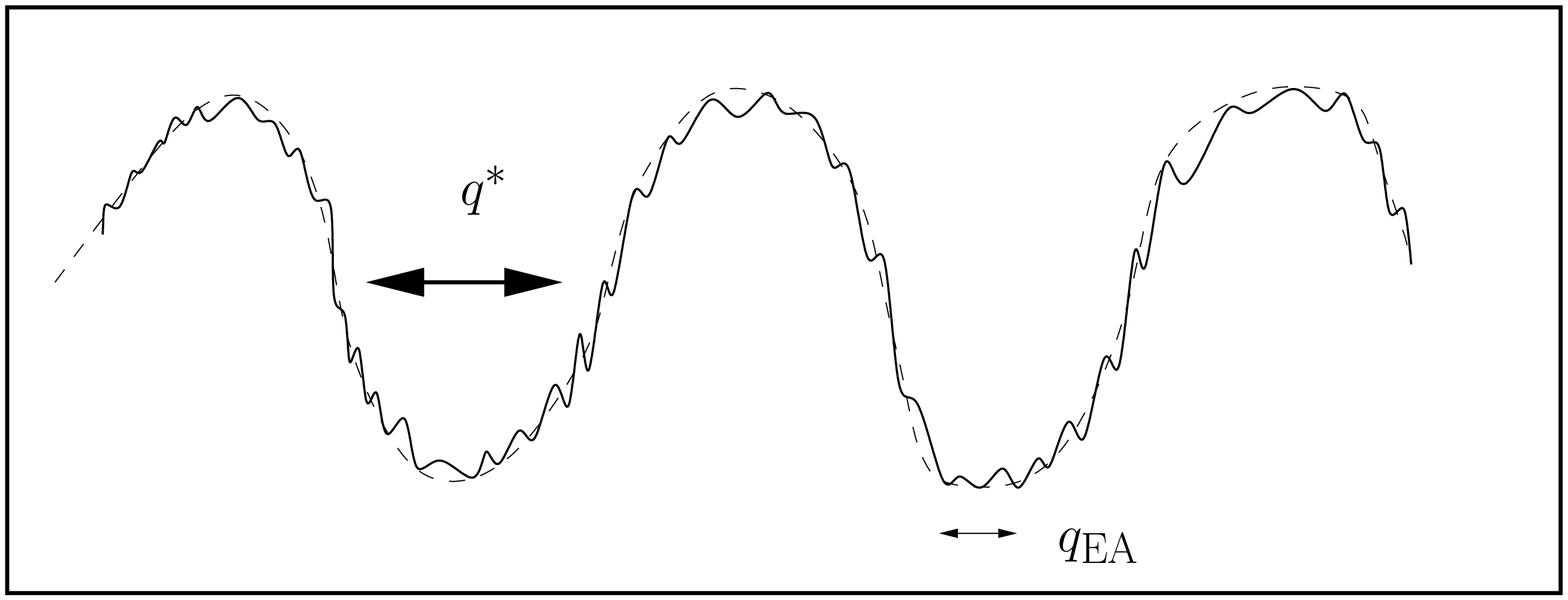}
\includegraphics[width=.5\textwidth]{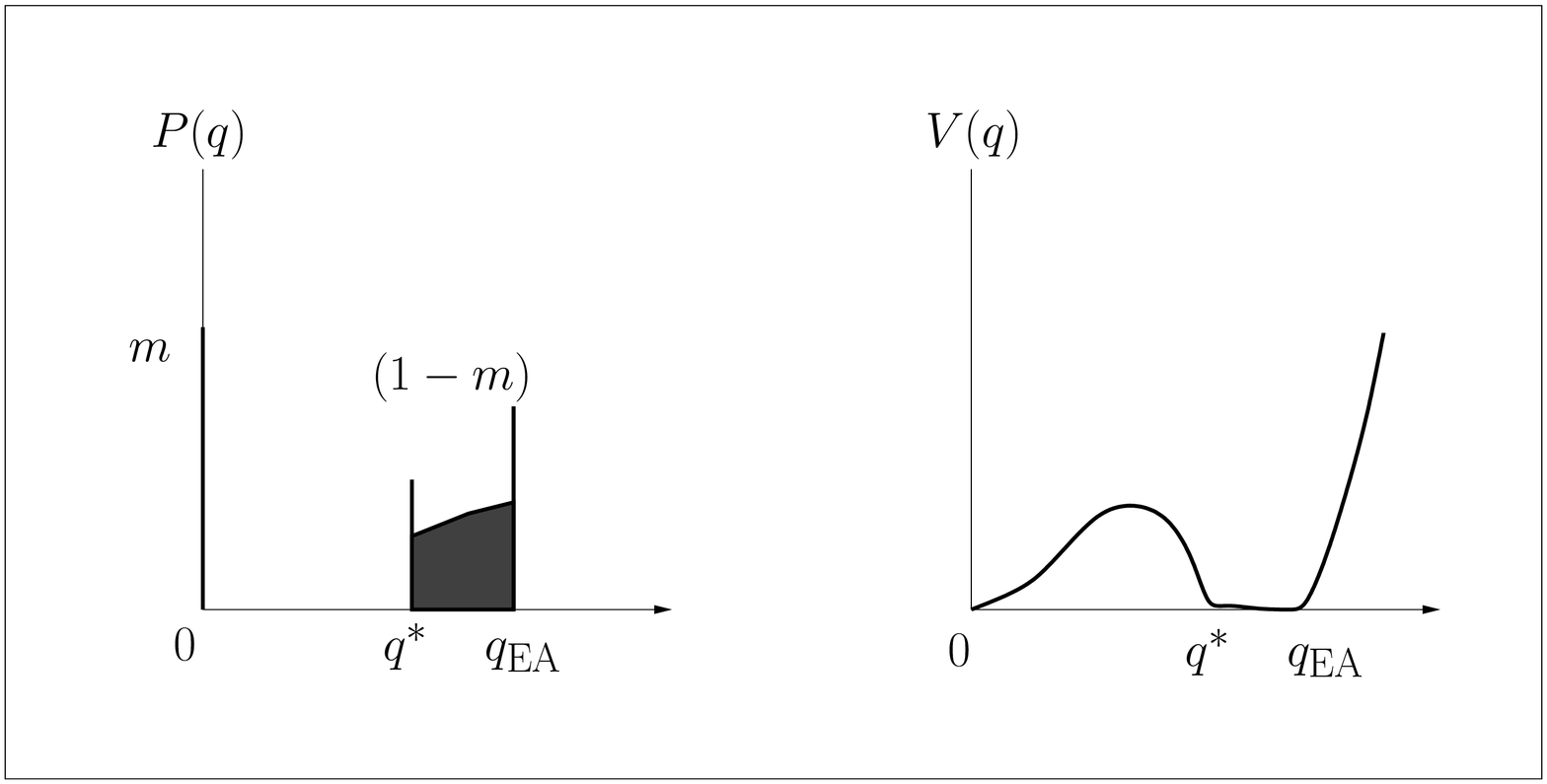}
\caption{
A sketch below the Gardner transition at equilibrium, $T \le T_{\rm G}$ or $P \ge P_{\rm G}$.
Top: a cartoon of the free energy. 
Bottom: effective potential $V(q)$ and Parisi distribution $P(q)$ beyond the Gardner point. 
The parameter $m$ is related to the probability of being 
in some state with  $q^*<q<q_{\rm EA}$, which is given by $1-m$.
}
\label{pq2}
\label{well2}
\end{figure}

Let us now consider higher pressures, or lower temperatures.
In most systems, there is a second transition discovered by Gardner~\cite{Ga85,MR03,MR04} years ago,  
at which each state itself breaks into smaller substates. The sketch one usually makes is as in Fig.~\ref{well2}.
To be more precise, we consider what happens with the effective potential and $P(q)$ for an equilibrium configuration
beyond  the Gardner point. The situation is depicted in Fig \ref{pq2}: there are many configurations at all distances
between $q^*$ and $q_{\rm EA}$: the state of size $q^*$ has fractured into many subcomponents of smaller sizes. 
However, going away from a configuration, up to correlations smaller than  $q^*$, one finds big barriers and no states,
up until completely different states, having minimal overlap are reached. The Parisi parameter $m$ is now related to the probability of being 
in some state closer than $q^*$, i.e. within the ``metabasin'' \cite{heuer}: this probability is given by $1-m$.
This fracturing of a state into many smaller ones also happens at the level of metastable states \cite{BFP97,MR03}: 
there is a line in the phase diagram where all states undergo a Gardner transition  (Fig.~\ref{pspi}).
Metastable states may be found as above~\cite{MR03}, by considering $m$ as a free parameter.
However, in the more complex regime 
beyond the Gardner transition it is not clear how to compute the threshold level that will dominate the dynamics~\cite{MR03,MR04}. 
It is possible that the threshold level could be identified by looking at the stability 
properties of the fluctuations {\em at the level of $q^*$}, but this need to be clarified.
See~\cite{Rizzo} for some initial steps in this direction.

\subsection{Vibrational modes and  dynamics close to jamming}

\begin{figure}[t]
\includegraphics[width=.5\textwidth]{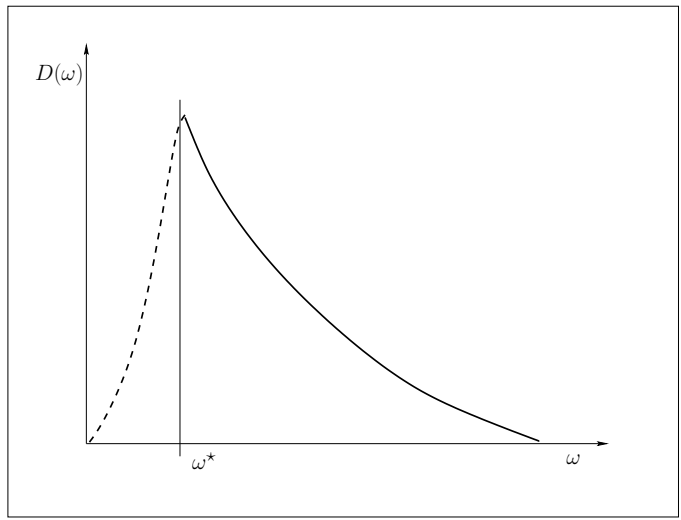}
\caption{
A schematic picture of the spectrum close to jamming.
}
\label{spectr}
\end{figure}

A system of hard spheres when compressed suddenly ends up in a configuration that is blocked, with the exception of
a small percentage of ``rattlers'' that are free to move within a ``cage'' made by their  neighbors. 
A mechanically stable  system of hard constituents such as this may be hypostatic, hyperstatic or isostatic, depending on whether the
number of contacts is less, more, or precisely just what it takes to guarantee mechanical stability. 
Because the system is prepared with a rapid compression, it
seems unlikely that it will be hyperstatic, because   if at some time during the compression it reaches stability, it is unable to move 
on to create further,
redundant contacts. The hypothesis that is usually made is that, forgetting the rattlers, the rest of the system is precisely 
isostatic, the assumption being  that there are no ``rattling clusters'' 
other than isolated rattling particles. 
For a detailed discussion of the
fundamental role of isostaticity in jammed packings see Refs.~\cite{LNSW10,He10}.
An additional assumption that seems to be justified in practice is that isostaticity is  ``irreducible'', in the sense that there is no subset  
of particles that is separately isostatic: if
 in such a system a contact is broken, then  by  definition all the particles in the system eventually become mobile.    
Clearly, this is a critical situation. 
Indeed, it has been proposed in Ref.~\cite{WSNW05,BW09b,LNSW10} that
these packings are marginally stable from a mechanical point of view, and from this 
most of the anomalous scalings that are found numerically have been derived analytically.
In particular, the criticality manifests itself in the spectrum of vibrations $D(\omega)$
at densities just below jamming~\cite{BW06,BW07},
which has a general shape as in Fig.~\ref{spectr}, where one has to distinguish two features:
\begin{itemize}
\item There is a branch of higher frequency modes, whose lowest frequency is $\omega^*$. The frequency $\omega^*$ goes
to zero as the pressure goes to infinity.
\item Within the  gap $0<\omega<\omega^*$ there are the acoustic modes, which exist even at finite pressures. 
Moreover, in Refs.~\cite{WSNW05,BW06,BW07,LP09,XVLN10,ML11} it was shown that the softer modes do not look like plane waves, therefore
acoustic modes are mixed with other kinds of soft modes.
\end{itemize}
In the rest of this section we will argue that the Gardner transition provides a natural explanation for the presence of soft modes at 
$\omega<\omega^*$.
These modes should appear
at all pressures beyond the Gardner transition.
However, the connection between the soft modes observed in Refs.~\cite{WSNW05,BW06,BW07,LP09,XVLN10,ML11} and the ones associated to the Gardner transition is not clear for the moment.

Consider the squared displacements $\widehat\Delta_i(t,t') = |x_i(t) - x_i(t')|^2$, where $i$ labels the $N$ particles of system,
and its average over particles $\wh\Delta(t,t') = N^{-1} \sum_{i=1}^N \wh\Delta_i(t,t')$.
In the following we will assume that the system has been prepared by some rapid compression at time $t=0$, in such a way that if $t>t'>0$
and $t'$ is large enough, the system is stuck into a glass state. 
The mean square displacement is given by the average of the squared displacement over the dynamical process,
\beq
\Delta(t,t') = \langle \wh\Delta(t,t') \rangle =  \frac1N \sum_{i=1}^{N} \langle |x_i(t) - x_i(t')|^2 \rangle
\eeq
and the variance of the squared displacement defines the so-called four-point susceptibility
\beq\begin{split}
\chi_4(t,t') &= N \left[ \left\langle \wh\Delta(t,t')^2 \right\rangle
-  \left\langle \wh\Delta(t,t') \right\rangle^2 \right] \\
&= \frac1N
\sum_{ij}  \left[
\langle |x_i(t) - x_i(t')|^2 |x_j(t) - x_j(t')|^2 \rangle
- \langle |x_i(t) - x_i(t')|^2 \rangle \langle |x_j(t) - x_j(t')|^2 \rangle
\right]
\end{split}\eeq
These definitions can be made more precise to take into account the presence of rattlers, we refer the reader to Ref.~\cite{IBB12} for
a detailed discussion. The ``cage size'' is the limit
\begin{equation}
\Delta^2(\infty)=  \lim_{t-t'\to\io } \lim_{ t' \rightarrow \infty} \Delta(t,t') 
\label{D2}
\end{equation}
where `$\infty$' stands for times $t,t'$ as large as the lifetime of the state.
At pressure $P$, the natural scale of the displacements is $1/P$, therefore it is convenient to introduce a scaled cage size as
\beq\label{Dio}
\Delta_\io = P^2 \Delta^2(\infty)  \propto \int_0^\infty d\omega \; \frac{D(\omega)}{\omega^2}  
\eeq
and the last relation is derived in Refs.~\cite{BW09b,IBB12}.
For $d>2$, this quantity is finite for finite $P$, because $D(\omega) \sim \omega^{d-1}$ in the low frequency acoustic branch, 
however it diverges as $P \rightarrow  \infty$ because $\omega^*$ goes to zero and
the integral is dominated by $  \int_{\omega^*}^\infty d\omega \; \frac{D(\omega)}{\omega^2}  \sim \frac{D(\omega^*)}{\omega^{*}}$
(while the integral in $0<\omega<\omega^*$ does not contribute to the divergence).
The fluctuations of the cage size yield the four-point susceptibility (or ``spin glass susceptibility'')~\cite{IBB12}
\begin{equation}
\chi_4(\infty) = \lim_{t -t'\rightarrow \infty} \lim_{ t' \rightarrow \infty}  \chi_4(t,t')  \sim  \int_0^\infty d\omega \; \frac{D(\omega)}{\omega^4}  \ .
\label{x4}
\end{equation}
On the one hand, from the theory of the Gardner transition, we expect $\chi_4(\infty)$ to diverge there, and to stay infinite up to infinite pressure. 
In fact, one may convince oneself that this is so just by considering the curvature of the effective potential above and below
the Gardner transition, where $\frac{d^2 V(q)}{dq^2}=0$. On the other hand, we may look at this from the point of view of normal modes:
 we  split   (\ref{x4}) in a contribution  above, and one below   $\omega^*$:
\begin{equation}
\chi_4(\infty) = \int_0^{\omega^*} d \omega \; \frac{D(\omega)}{\omega^4}  +
 \frac{D(\omega^*)}{(\omega^{*})^3} 
\label{x4_1}
\end{equation}
It has already been remarked in Ref.~\cite{IBB12} that in three dimensions even
the acoustic modes will make the integral in $0<\omega<\omega^*$ diverge {\em for finite $P$}. 
However, the effect of acoustic modes shows up
in $\chi_4(t,t')$ only at
very long time differences $t-t' \gg 1/\omega^*$, so that in Ref.~\cite{IBB12} 
it was shown that the regime of $t-t' \sim 1/\omega^*$ gives
a good definition of the four-point susceptibility.
In our large-dimensional case (actually, for all $d>4$), 
the density of acoustic modes is negligible, but we still expect that the first term in (\ref{x4_1}) diverges below the Gardner transition.
The conclusion seems to be that there are other soft modes (below $\omega^*$) that do not contribute to the  linear susceptibilities
or to the short-time $t-t'$ value of $\chi_4(t,t')$, but dominate the limit of $\lim_{t-t' \rightarrow \infty} \chi_4(t,t')$. 
Although it is tempting to identify these modes with the ones observed in Refs.~\cite{WSNW05,BW06,BW07,LP09,XVLN10,ML11}, more work is
needed to clarify the connection.

\subsection{Out of equilibrium dynamics in the Gardner phase}

\begin{figure}
\includegraphics[width=.5\textwidth]{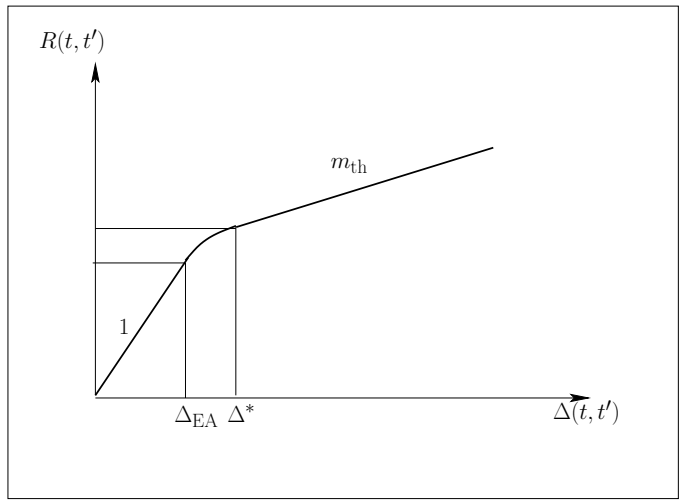}
\caption{
A sketch of an FDT plot for an aging  system below the Gardner transition. The plot is composed of two straight
and a curved segment.
}
\label{fdt}
\end{figure}

The out of equilibrium dynamics of this system has not been solved, but from the structure of states one may already
guess its main features. Below the Gardner transition line, the slow compaction (aging) dynamics  should proceed
close to the threshold level, defined as described above as the one where the stability at the level of $q^*$ is marginal.
The relaxation process can be seen as a dynamical exploration of phase space starting from a completely correlated state ($q=1$) down
to a completely decorrelated state ($q=0$).
The relaxation should be fast from correlation $q=1$ down to $q_{\rm EA}$, and then proceed  -- in a progressively  slower way as the system ages --
down to $q^*$, and from there to zero.
The fluctuation-dissipation properties may be studied by considering a system with hard spheres in a thermal bath of temperature 
unity, subjected to a pressure $P$ generated by either a piston or by coupling to the radii of all spheres.
The response and correlation functions are as described in Refs.~\cite{Pa97,BB02}: the response is computed from the 
staggered displacement  $R(t,t') = \sum_i \xi_i \delta \langle x_i \rangle $ induced subjecting  particles to random unit fields
$\xi_i$  with an energy term $E_{field} = h \sum_i \xi_i \delta x_i$, per unit of $h$. The conjugate correlation may be taken to
be the quadratic displacement $\Delta(t,t')$ defined above.
Response and correlations may be put together in a plot, which should look as in Fig.~\ref{fdt} for long times $t'$, the time
$t>t'$ being used to produce a parametric plot of $R$ versus $\D$. Here,
$\Delta^*$ and $\Delta_{\rm EA}$ are the values corresponding to the correlations $q^*$ and $q_{\rm EA}$. 
At every time, the first barriers encountered are the small ones close to $q_{\rm EA}$: beyond the Gardner transition, 
where small states are separated by relatively small excitations because there are states at all distances $q$ with  $q^*<q<q_{\rm EA}$.
This might help explain the paradox that the path between these small states is mainly along the flattest  vibrational modes --
as found in Ref.~\cite{BW09} -- while this is not what one expects for the large-scale relaxation within a supercooled
liquid, at least within the RFOT scenario.

\subsection{Low temperature excitations}

A long standing problem in the physics of glassy and amorphous materials is the  low temperature behavior of their specific heat and thermal conductivity, which turns out to be quite different to that observed in crystals. These 
features are all the more intriguing because they tend to be quite universal  for all amorphous materials.
 The usual explanation for this phenomenon is to attribute it to localized quantum-mechanical two-level tunneling 
 systems~\cite{AHV72,ph81}.
 These models assume that there are particles or groups of particles that evolve and tunnel 
 in random local potentials. These potentials are usually proposed phenomenologically, although they are of course generated
by the same interactions that produced the amorphous solid in the first place.
Furthermore, these simple localized clusters or particles will be  coupled, and their interactions might generate collective effects.
A proposal to take these features into account~\cite{kuhn} is to consider a system of strongly coupled localized deformations, 
which one may assimilate as ``spin''-like excitations,  and to assume that they have essentially random 
interactions -- with long range, partly because elasticity is long range, and partly to make  the system solvable.
One obtains in this way a ``spin-glass'' of deformations, with elementary excitations which one may calculate and which tend to
be universal because of their collective nature.  Note that this way, phenomenology has been pushed one step up, to the
effective interaction of excitations.

Quite clearly, the mechanisms that generate the coupling between low-temperature excitations, and the one responsible 
for the amorphous matrix on which they live, are one and the same. One would thus expect the same theory to explain both
features. In the context of this paper, it is very tempting to interpret the large valleys (metabasins) of size $q^*$ as being
the amorphous structure, and the excitations of all sizes between $q^*$ and $q_{\rm EA}$ as a ``spin-glass'' of small excitations
within that amorphous structure. Formally this is clearly so, a fact that was already recognized by Gardner in her original paper, 
where she showed that the transition is essentially that of the Sherrington-Kirkpatrick model within each large
state. More recently, the analogy between jammed packings and the Sherrington-Kirkpatrick model has also been underlined~\cite{Wy12},
and the idea that there is a deep connection between the marginality of jammed packings and low-temperature anomalies in glasses
was also proposed by S.~Nagel (e.g. in his talk at the ACS meeting, Philadelphia, August 2012).
A possible difference may be noted with respect to Ref.~\cite{kuhn}: here the spin-glass transition need not 
(and in general will not) coincide with the liquid-glass transition at which the amorphous matrix is formed.

\section{Replicated entropy of infinite-dimensional hard spheres}
\label{sec:II}

The above discussion provides several important motivations to look for an instability of the 1RSB solution in particle
systems, akin to the Gardner transition of spin glasses~\cite{Ga85}. 
Additional ones will be given by the more technical discussion that we now start, see Sec.~\ref{sec:tech1RSB}.
We will show that a Gardner instability indeed happens in hard sphere
systems in the limit $d\to\io$ (and probably also in finite dimensions within the mean-field RFOT scenario).

We will consider a system of $N$
hard $d$-dimensional spheres with unit diameter, enclosed in a volume $V$, hence at density $\r=N/V$. The packing fraction is $\f = 2^{-d} \r V_d$, 
with $V_d = \pi^{d/2}/\G(1+d/2)$ the volume of a sphere of unit radius.
In the first paper of this series~\cite{KPZ12}, 
we derived an exact expression for the replicated entropy
of this system in the limit of large dimension $d\to\io$. 
The result is obtained by first writing the  entropy in
a manifestly rotationally and translationally invariant form, and then performing a saddle point evaluation
in the limit $d\to\io$. Within replica theory the resulting entropy is a function of the  density function $\r(\hat q)$,
where the matrix $\hat q$ is a $m\times m$ symmetric 
matrix that encodes the overlaps $q_{ab}$ between different replicas.
The main result of Ref.~\cite{KPZ12} was that a Gaussian assumption for
$\r(\hat q)$ gives the exact result for the entropy, i.e. for all thermodynamic properties of the system, exclusively
in terms of $q_{ab}$ (with $\sum_{a=1}^m q_{ab}=0$ for all $b$, because of translational invariance). 
In other words, no higher order parameters $q_{abc},q_{abcd}...$ are necessary
in the large dimensional limit. 

The proof of Ref.~\cite{KPZ12} was restricted to the 1RSB form of $q_{ab}$.
In this section, we will extend the results of Ref.~\cite{KPZ12} to
obtain the replicated entropy as a function of the overlap matrix, without making any assumption on the RSB structure.
We will start by deriving the Gaussian replicated entropy for a generic overlap matrix, and then show that this form
coincides with the exact result. Obviously, in this section we will often make reference to Ref.~\cite{KPZ12}, which we encourage
the reader to consult before looking to the rest of the section. Another option is to skip this section and take the result
-- as expressed by Eqs.~\eqref{eq:gauss_r} and \eqref{eq:gauss_a} -- for granted. This will be the starting point to study 
the stability of the 1RSB solution (and much more) in the following.

\subsection{Gaussian ansatz for a generic overlap matrix}


We have to parametrize a generic Gaussian form of $\r(\hat q)$, or equivalently $\r(\bar u)$, where $q_{ab} = u_a \cdot u_b$
and $u_a$ are the $d$-dimensional vectors corresponding to replica displacements, with $\bar u = \{u_1,\cdots,u_m\}$.
We can choose a parametrization in terms of a $m \times m$ symmetric matrix $\hat A$ such that $\sum_{a=1}^m A_{ab} = 0$
for all $b$. Calling $\hat A^{m,m}$ the $(m-1) \times (m-1)$ matrix obtained from $\hat A$ by removing the last line and column,
the most general Gaussian form of $\r(\bar u)$ is
\beq\label{Gaussparam}
\r(\bar u) = \frac{\r \, m^{-d}}{(2 \pi)^{(m-1)d/2} \det(\hat A^{m,m})^{d/2}} e^{-\frac12 \sum_{ab}^{1,m-1} (\hat A^{m,m})^{-1}_{ab} u_a \cdot u_b}
\eeq
which is normalized according to $\r = \int \DD \bar u \r(\bar u)$
and $\DD \bar u = m^d \d(\sum_a u_a) du_1 \cdots du_m$.
The parameters $A_{ab}$ are interpreted as
\beq\label{eq:alpha}
\la u_a \cdot u_b \ra = \frac{1}\r \int \DD \bar u \r(\bar u) u_a \cdot u_b = d \, A_{ab} \ , 
\eeq
for $a,b \in [1,m-1]$, while $\la u_a \cdot u_m \ra = -\sum_{b=1}^{m-1} \la u_a \cdot u_b \ra = A_{am}$ and
$\la u_m \cdot u_m \ra = \sum_{ab}^{1,m-1} \la u_a \cdot u_b \ra = A_{mm}$.


The saddle point value of $\hat q$, that dominates all the integrals over $\r(\hat q)$, 
is obtained as follows.
We start from the normalization condition (note that a complete derivation of $J(\hat q)$, that was not reported
in Ref.~\cite{KPZ12}, is reported here in Appendix~\ref{app:J})
\beq
\r = \int d\hat q J(\hat q) \r(\hat q)
\propto \int d\hat q
 \prod_{a=1}^m \d\left( \sum_{b=1}^m q_{ab} \right)
\ \ e^{\frac12 (d - m) \log \det \hat q^{m,m} - \frac12 \sum_{ab}^{1,m-1} (\hat A^{m,m})^{-1}_{ab} q_{ab}  } \ ,
\eeq
and maximizing the exponent for $d\to\io$ leads, for $a,b \in [1,m-1]$, to $(q^{m,m})^{-1}_{ab} = (\hat A^{m,m})^{-1}_{ab}/d$, hence
$q^{sp}_{ab} = d \, A_{ab}$, consistently with Eq.~\eqref{eq:alpha}.

To compute the replicated entropy using the general Gaussian ansatz we start from Eq.~(45) of Ref.~\cite{KPZ12},
which gives the following expression for the replicated entropy:
\beq\label{eq:gauss_sp}
\begin{split}
\SS[\r(\hat q)]/N & =  1 - \log \r(\hat q^{sp}) -
2^{d-1} \f \,
 \FF\left(\frac{d}{D^2}  2 \hat q^{sp} \right) \ ,
\end{split}\eeq
where the function $\FF$ is given in Eq.~(37) of Ref.~\cite{KPZ12}.
The ideal gas term is
\beq
 1 - \log \r(\hat q^{sp}) = 1 - \log\r + d \log m + \frac{(m-1)d}{2} + \frac{(m-1)d}{2} \log(2 \pi) + \frac{d}2 \log \det(\hat A^{m,m}) \ .
\eeq
The interaction term is
\beq
2^{d-1} \f \,
 \FF\left(\frac{d}{D^2}  2 \hat q^{sp} \right) = 2^{d-1} \f \,
 \FF\left(\frac{d^2}{D^2}  2 \hat A \right) \ .
\eeq
Therefore
\beq
\frac{\SS[\hat A]}N  = 1 - \log\r + d \log m + \frac{(m-1)d}{2} \log(2 \pi e) + \frac{d}2 \log \det(\hat A^{m,m}) -  2^{d-1} \f \,
 \FF\left(\frac{d^2}{D^2}  2 \hat A \right) \ .
\eeq


In order to obtain a simple limit $d\to\io$,
it is convenient to define a matrix $\hat \a = \frac{d^2}{D^2} \hat A$ and 
a reduced packing fraction
$\wh \f = 2^d \f / d$. With this choice we have
\beq\label{eq:gauss_r}
s[\hat \a] =\frac{\SS[\hat \a]}N  = 1 - \log\r + d \log m + \frac{(m-1)d}{2} \log(2 \pi e D^2/d^2) + \frac{d}2 \log \det(\hat \a^{m,m}) -  \frac{d}2 \wh \f \,
 \FF\left( 2 \hat \a \right) \ .
\eeq
The matrix $\hat\a$ is a variational parameter and is therefore determined by maximization of the entropy. Defining 
$\FF'_{ab}(\hat\y) = \frac{d\FF(\hat\y)}{d\y_{ab}}$, the equation for $\hat\a$ is
\beq\label{eq:gauss_a}
(\hat\a^{m,m})^{-1} = 2\wh\f \FF'_{ab}(2 \hat\a) \ .
\eeq
Eqs.~\eqref{eq:gauss_r} and \eqref{eq:gauss_a} provide the expression of the Gaussian replicated entropy and
will be the starting point of all our calculations.

Note that the entropy of the equilibrium glass is obtained by optimizing 
$s[\hat\a]/m$, given in
Eq.~\eqref{eq:gauss_r}, with respect to the matrix $\hat\a$ and of $m$.
Let us call $\hat\a^*$ and $m^*$ the optimal values, $\hat\a^*$ being the solution of Eq.~\eqref{eq:gauss_a}.
The reduced pressure $p = \b P/\r$ of the equilibrium glass is given by
\beq
p_{glass}(\f) = -\frac{\f}{m^*} \frac{\partial s[\hat \a^*]}{\partial \f}
= \frac{1}{m^*} \left[ 1 +
\frac{d}2 \wh \f \,
 \FF\left( 2 \hat \a^* \right) \right] \ .
\eeq
This result shows that the pressure diverges whenever $m^*\to 0$ as $p \sim 1/m^*$. Hence, the density at which $m^*\to 0$ 
defines the jamming point~\cite{PZ10}.

\subsection{Exact computation}

We now show that Eqs.~\eqref{eq:gauss_r} and \eqref{eq:gauss_a} can be equivalently obtained by an exact evaluation
of the saddle point equations derived in Eqs.~(64) and (65) of Ref.~\cite{KPZ12}.
In fact, we can make use of Eqs.~(65) and (39) of Ref.~\cite{KPZ12}
to obtain a closed self-consistent equation for $\hat q^{sp}$, which
as before is the point where the argument of the integral
\beq\label{exactapp1}
\r = \int d\hat q J(\hat q) \r(\hat q)
= e^\l m^d C_{m,d} \int d\hat q
 \prod_{a=1}^m \d\left( \sum_{b=1}^m q_{ab} \right)
\ \ e^{\frac{d}2 \log \det \hat q^{m,m}
- d \wh\f \FF\left(\frac{d}{D^2} (\hat q + \hat q^{sp}) \right)
}
\eeq
is maximum (subleading terms for $d\to\io$ have been neglected). Taking the derivative with respect to $\hat q$ and computing the result in
$\hat q = \hat q^{sp}$ leads to the equation
\beq
\frac{D^2}d (\hat q^{sp;m,m})^{-1} = 2 \wh\f \FF'_{ab}\left(2 \frac{d}{D^2} \hat q^{sp} \right) \ .
\eeq
Clearly, defining $\hat \a = \frac{d}{D^2} \hat q^{sp}$ this equation is equivalent to Eq.~\eqref{eq:gauss_a}.

Evaluation of Eq.~\eqref{exactapp1} at the saddle point gives the equation for $\l$
\beq
\r =  e^\l m^d C_{m,d} \  e^{\frac{d}2 \log \det \hat q^{sp;m,m}
- d \wh\f \FF\left(2 \frac{d}{D^2} \hat q^{sp} \right)
} \ ,
\eeq
from which, using Eq.~(78) of Ref.~\cite{KPZ12}, we obtain
\beq\label{exactapp2}
-\l = -\log\r + d\log m + \frac{d}2 (m-1) \log\left(\frac{2 \pi e}d\right) + \frac{d}2 \log \det \hat q^{sp;m,m}
- d \wh\f \FF\left(2 \frac{d}{D^2} \hat q^{sp} \right) \ .
\eeq
Combining Eqs.~(64) and (65) of Ref.~\cite{KPZ12}, using Eq.~\eqref{exactapp2} and recalling the definition $\hat \a = \frac{d}{D^2} \hat q^{sp}$
 we have
\beq
\begin{split}
\SS[\r(\hat q)]/N & =  1 - \l +
\frac{d}2 \wh\f \,
 \FF\left(\frac{d}{D^2}  2 \hat q^{sp} \right) \\
& = 
  1 -\log\r + d\log m + \frac{d}2 (m-1) \log\left(\frac{2 \pi e}d\right) + \frac{d}2 \log \det \hat q^{sp;m,m}
- \frac{d}2 \wh\f \,
 \FF\left(\frac{d}{D^2}  2 \hat q^{sp} \right) \\
& = 
  1 -\log\r + d\log m + \frac{d}2 (m-1) \log\left(\frac{2 \pi e D^2}{d^2} \right) + \frac{d}2 \log \det \hat \a^{m,m}
- \frac{d}2 \wh\f \,
 \FF\left( 2 \hat \a \right)
\end{split}\eeq
which coincides with Eq.~\eqref{eq:gauss_r}. This completes the proof of the exactness of the Gaussian ansatz for the computation
of the entropy. Note that as already observed in Ref.~\cite{KPZ12} this does not imply that the Gaussian form \eqref{Gaussparam} can
be used to compute correlation functions (that encode structural properties), because the equivalence is only correct at the saddle point level for the entropy: 
this is consistent with the numerical observation of a non-Gaussian cage shape obtained in Ref.~\cite{CIPZ12}.
A computation of the cage shape is in progress and will be hopefully reported in future papers of this series.

\section{1RSB solution}
\label{sec:IIa}

\subsection{Reminder of the 1RSB solution}

in Ref.~\cite{KPZ12,PZ10,PZ06a} we studied the 1-step replica symmetry breaking (1RSB) ansatz, 
which amounts in this formalism
to assuming that all replicas are equivalent~\cite{Mo95}. 
For completeness let us recall here this result, which
correponds to the simple choice
\beq\label{alpha1RSB}
\a_{ab}^{\rm 1RSB}= \wh A \left( \delta_{ab}-\frac1m \right) \ ,
\eeq
with $\wh A =  \frac{d^2}{D^2} A$. Within this ansatz, $A$ is the ``cage radius'', as it is proportional to the long
time limit of the mean square displacement in the glass~\cite{PZ10}.
Note that we use a small hat for matrices, while the wide hat just
denotes reduced scalar variables. Hence $\hat A$ is a matrix while $\wh A$ is a scalar, and they should not be confused.

Using the relation $\log\det\{ [\wh A (\d_{ab}-1/m)]^{m,m} \} = (m-1)\log \wh A - \log m$ 
and the results of Sec.~VIIB of Ref.~\cite{KPZ12},
it is easy to check that Eq.~\eqref{eq:gauss_r}
reduces to the
result of Ref.~\cite{KPZ12} for the 1RSB entropy, which is
\beq\label{pap1:1RSB}
\begin{split}
s[\hat\a^{\rm 1RSB}] &= 
1 - \log\r + \frac{d}{2} \log m + \frac{(m-1)d}{2} + \frac{(m-1)d}{2} \log\left(\frac{2 \pi D^2 \wh A}{d^2}\right) 
- \frac{d}2 \wh \f  [1 - \GG_m(\wh A) ] \ , \\
\GG_m(\wh A) &= 1 - m  \int_{-\io}^\io \frac{d\l}{\sqrt{2 \pi}} e^{-\frac12 \l^2 }
\left[ \frac12 \left( 1 + \erf\left( \frac{ \sqrt{2 \wh A} - \l }{\sqrt{2} }  \right) \right) \right]^{m-1}
 \ .
\end{split}\eeq
The equation for $\wh A$ is derived by optimizing the above results, leading to
\beq\label{pap1:1RSBeq}
\frac{m-1}{\wh A} + \wh \f \frac{d \GG_m(\wh A)}{d \wh A}=0 
\hskip5pt
\Rightarrow
\hskip5pt
\frac1{\wh\f} = \frac{\wh A}{1-m}  \frac{d \GG_m(\wh A)}{d \wh A} = \FF_m(\wh A) \ .
\eeq
The function $\FF_m(\wh A)$ introduced
here should not be confused with the function $\FF(\hat\a)$ introduced before.
These are different functions, the first acts on a scalar while the second on a matrix.

The physical consequences of this expression for the entropy have been derived in Ref.~\cite{PZ06a,PZ10}, where the expressions
of the dynamical transition density, the Kauzmann transition density, and the GCP density have been derived, with the scalings
sketched in Fig.~\ref{pspi}.
Furthermore,
at the level of the 1RSB solution, we know that $\wh A^* \sim m^*$ so we conclude that the cage radius vanishes as
$\wh A^* \sim 1/p$. As a consequence of this scaling, the scaled cage size $\D_\io$ introduced in Eq.~\eqref{Dio} is found to
diverge as $\D_\io \sim p^2 \wh A^* \sim p$, as noted in Ref.~\cite{IBB12}.

\subsection{Inconsistencies of the 1RSB solution}
\label{sec:tech1RSB}

The 1RSB predictions for physical
quantities were carefully compared with numerical results, around both the glass and the 
jamming transitions~\cite{PZ10,BJZ11,CIPZ11,CIPZ12,CCPZ12}. 
Despite the good overall agreement with numerical data, one expects, as described above, that a Gardner transition
to a full replica symmetry breaking scheme is generic.
Furthermore, 
several inconsistencies have been found
close to the jamming transition, at very high pressure:
\begin{itemize}
\item
The 1RSB solution predicts the existence of jammed packings with density $\f_j$ in the interval
$\frac{\f_j}{2^{-d} d} =\wh \f_j \in  [6.26, \log d]$~\cite{PZ10}.
However, only the packings with $\wh\f_j \sim \log d$ are isostatic, with each particle
 in contact, on average, with $z=2d$ other particles.
Instead, the packings with $\wh \f_j$ of order 1 are found to be hyperstatic with $z>2d$, which is, as mentioned above,
 unexpected and inconsistent with numerical
results. 
\item 
In the glass phase, the exact relation between the pressure $p$ and the contact value $y(\f)$ of the 
pair correlation, $p = 1 + 2^{d-1} \f y(\f)$~\cite{Hansen}, is violated. In particular, it is found
that when $\f\to\f_j$, $p\sim d \, \f_j/(\f_j-\f)$, consistently with numerical results, while
\beq
y(\f) = \frac{ d  }{2^{d-1} (\f_j - \f) } \times \frac{ 1 }{ 1 - 2^{1-d} \, d/\f_j } \ ,
\eeq
where the first term is the one that is consistent with the scaling of the pressure.
Hence, the correct relation between $p$ and $y(\f)$ is 
recovered only if $2^{1-d} \, d/\f_j = 2 / \wh \f_j \ll 1$ when $d\to\io$,
which again suggests that the 1RSB solution is inconsistent when $\wh \f$ is of order 1 and might
be stable only when $\wh\f \gg 1$.
\item The scaling at large (reduced) pressure $p$ of the cage radius $A$ (the long time limit of the
mean square displacement in the glass) 
predicted by the 1RSB
solution is $A \sim p^{-1}$, 
while the marginal stability argument of Ref.~\cite{WSNW05,BW09b} predicts
that $A \sim p^{-3/2}$, which has been confirmed numerically in several studies, e.g. Refs.~\cite{WSNW05,BW09b,IBB12}.
This exponent controls all the other exponents that characterize criticality 
at the jamming transition~\cite{IBB12} and is directly related to the 
anomalous soft vibrational modes that appear at jamming~\cite{LNSW10,He10,BW09b,IBB12}, 
hence reconciling the theoretical prediction with the numerical results is of extreme importance.
\item Other exponents that characterize the structure at jamming, for instance the famous (almost)
square-root singularity in the pair correlation function~\cite{OLLN02,DTS05,SLN06,CCPZ12}, are not
reproduced by the 1RSB solution, at least at the Gaussian level (a more detailed calculation of the structure functions
based on the non-Gaussian theory developed in this series of papers is in progress
and will hopefully be reported in a future paper).
\end{itemize}

All these considerations  suggest strongly that  the 1RSB solution is unstable, at least when pressure is large enough
and $\wh \f$ is of order $1$. They provide further motivations to
 study the stability of the 1RSB solution~\cite{Ga85}, which is the subject of the next section.

\section{Second order expansion: the Hessian matrix}
\label{sec:III}

Here we obtain the expansion around the 1RSB solution at the quadratic order.
For this we need to consider a more general ansatz or
expand around the 1RSB solution, which is made possible by the
general expression of the entropy obtained in Eqs.~\eqref{eq:gauss_r} and \eqref{eq:gauss_a}.
The quadratic expansion of the entropy around the 1RSB solution provides a stability matrix
whose eigenvalues allow one to determine the stability of the solution.
We will find that, as it happens in a Gardner transition~\cite{Ga85,MR03,MR04}, 
the 1RSB solution becomes unstable when pressure is large enough.
In finite and arbitrarily large dimensions, this happens for all $\wh\f_j$.
However, for $d\to\io$, the so-called {\it glass close packing} (GCP)~\cite{PZ10} which is
the densest amorphous packing and has $\wh\f \sim \log d$ becomes stable again, suggesting that 
the 1RSB predictions for jamming are still approximately useful as a starting point, but should be
corrected to take into account its instability.

\subsection{General structure of the Hessian matrix}

Because the matrix $\hat\a$ should have the sum of the elements of every column 
and every row equal to zero, we can say that the independent entries are the elements 
above the diagonal of the matrix, provided that the matrix is symmetric and the diagonal 
is fixed in such a way that the constraints on the sum over the elements in a row or in column is satisfied. 
Hence, in the following we denote as $\d/\d \a_{a<b}$ the derivative taken with respect
of the element $\a_{ab}$ with $a<b$ which is assumed to be the only independent element
(hence its variation induces a variation of $\a_{ba}$ and of the diagonal elements $\a_{aa}$ and
$\a_{bb}$).
Taking into account all this, we define
the Hessian matrix as
\beq
\begin{split}
M_{a\neq b,c\neq d}=\frac{2\wh A^2}d 
\left.\frac{\delta^2 s[\hat\a]}{\delta \a_{a<b}\delta\a_{c<d}}\right|_{\hat\a^{\rm 1RSB}}
=M_1\left(\frac{\delta_{ac}\delta_{bd}+\delta_{ad}\delta_{bc}}{2}\right)+M_2\left(\frac{\delta_{ac}+\delta_{ad}+\delta_{bc}+\delta_{bd}}{4}\right)+M_3
\end{split}
\eeq
where the replica structure is a consequence of the structure of $\hat\a^{\rm 1RSB}$.
Although the matrix $M$ is defined by the above equation only for $a<b$ and $c<d$, we will
define it for convenience also for $a>b$ and $c>d$ assuming that it is symmetric (hence
the notation $M_{a\neq b,c\neq d}$).
The prefactor $2 \wh A^2/d$ is chosen for later convenience and is positive, hence it
does not affect the sign of
the three different eigenvalues of the mass matrix, which are
\beq
\begin{split}
\l_R &= M_1 \ , \\
\l_L & = M_1 + (m-1) (M_2 + m M_3) \ , \\
\l_A &= M_1 + \frac12 (m-2) M_2 \ . 
\end{split}
\eeq
Defining the ``entropic'' and ``interaction'' terms
\beq\begin{split}
M^{(I)}_{a\neq b;c \neq d} & = \left. \frac{\delta^2 \mathcal F[\hat\y]}{\delta \y_{a<b}\delta \y_{c<d}} 
\right|_{\hat \y = 2 \hat \a^{\rm 1RSB}} \\
M_{a\neq b;c\neq d}^{(E)}&=\left.\frac{\delta^2}{\delta \a_{a<b}\delta\a_{c<d}}\log\det( \hat\alpha^{m,m} ) 
\right|_{\hat \a = \hat \a^{\rm 1RSB}}
\end{split}\eeq
we have
\beq
M_i =\wh A^2 M_i^{(E)}-4 \wh \varphi\, \wh A^2 M_i^{(I)} \ .
\eeq
We now compute these two terms separately.

\subsection{The entropic term}
\label{Mentropic}

To compute the entropic term it is convenient to introduce a shorthand notation $\hat\b = \hat\a^{m,m}$.
Let us also use indices $i,j,k,\cdots$ for $\b$ to highlight that they run from $1$ to $m-1$.
In the 1RSB solution $\b^{\rm 1RSB}_{ij} = \wh A ( \d_{ij} - 1/m)$ has the same form of $\hat \a$ but on the
reduced $(m-1)\times (m-1)$ space.
Then we have
\beq
(\hat\b^{\rm 1RSB})^{-1}_{ij} = \frac{1}{\wh A}(\delta_{ij}+1) \ .
\eeq
Moreover we have
\beq
\frac{\delta \b_{ij} }{\delta \a_{a<b}}=\delta_{ia}\delta_{jb}+\delta_{ib}\delta_{ja}-\delta_{ij}(\delta_{ia}+\delta_{ib}) \ .
\eeq
Using the standard formula
\beq
\frac{\d \log \det ( \hat \b)}{\delta \b_{ij} } = \b^{-1}_{ji}
\eeq
we have
\beq
\frac{\delta}{\delta \a_{a<b}}\log\det( \hat\b ) = \sum_{ij} \b^{-1}_{ji}  \frac{\delta \b_{ij} }{\delta \a_{a<b}} = \Tr \left[ \hat\b^{-1} \frac{\d \hat\b}{\d\a_{a<b}} \right] \ .
\eeq
From the definition $\hat\b \hat\b^{-1} = I$ where $I$ is the identity matrix we have
\beq
0 = \frac{\d}{\d \a_{c<d}}( \hat\b \hat\b^{-1})
\hskip10pt
\Rightarrow
\hskip10pt
\frac{\d \hat\b^{-1}}{\d \a_{c<d}} = - \hat\b^{-1} \frac{\d \hat\b }{\d \a_{c<d}} \hat\b^{-1} \ .
\eeq
It follows that
\beq
\begin{split}
M_{a<b;c<d}^{(E)}&=\left.\frac{\delta^2}{\delta \a_{a<b}\delta\a_{c<d}}\log\det( \hat\b ) 
\right|_{\hat \b = \hat \b^{\rm 1RSB}} =
- \Tr \left[ 
\hat\b^{-1}  \frac{\delta \hat\b}{\delta \a_{a<b}}  \hat\b^{-1} \frac{\delta \hat\b}{\delta \a_{c<d}}
\right]_{\hat \b = \hat \b^{\rm 1RSB}}
\\
&=- \frac{1}{\wh A^2}\left[2\left(\delta_{ac}\delta_{bd}+
\delta_{ad}\delta_{bc}\right)+\left(\delta_{ac}+\delta_{ad}+\delta_{bc}+\delta_{bd}\right)\right]
\end{split}
\eeq
from which it follows that
\beq\begin{split}
M_1^{(E)}&=M_2^{(E)}=- \frac{4}{\wh A^2} \ , \\
M_3^{(E)}&=0 \ .
\end{split}\eeq

\subsection{The interaction term}

We now consider the interaction term. Hence we need an expansion of the function $\FF(\hat\y)$ 
around the 1RSB solution.
Recall that $\hat\y$ is a $m\times m$ symmetric matrix such that the sum of the elements in each row and in each column 
is zero.
Starting from the results of~Section~V of Ref.~\cite{KPZ12},
we can write explicitly the function $\FF(\hat\y)$, introducing $m$-dimensional vectors $x_a$ such that
$x_a \cdot x_b = \y_{ab}$, as
\beq\label{FFreplica}
\begin{split}
\mathcal F[\hat \y] &= \int \frac{\de^m \epsilon}{(\sqrt{2\pi})^m} \exp\left[ -\frac 12 \min_a |\epsilon +x_a|^2    \right]=\\
&=\lim_{n\to 0}\sum_{n_1,\ldots, n_m; \sum_a^mn_a=n}\frac{n!}{n_1!\ldots n_m!}\exp\left[ -\frac 12 \sum_{a=1}^m \frac{n_a}{n}|x_a|^2+\frac{1}{2} \sum_{a,b}^{1,m}\frac{n_an_b}{n^2}x_a\cdot x_b  \right] \\
&=\lim_{n\to 0}\sum_{n_1,\ldots, n_m; \sum_a^mn_a=n}\frac{n!}{n_1!\ldots n_m!}\exp\left[ -\frac 12 \sum_{a=1}^m \frac{n_a}{n}\y_{aa}+\frac{1}{2} \sum_{a,b}^{1,m}\frac{n_an_b}{n^2}\y_{ab}  \right]
\end{split}\eeq

\subsubsection{1RSB value of $\FF$}

First let us compute $\FF$ on the 1RSB solution where
\beq
\y_{ab}^{\rm 1RSB}=2\wh A \left( \delta_{ab}-\frac 1m\right)\:.
\eeq
Let us call $\d_{a,{\rm min}}$ a function that is equal to one only if $a$ is such that $\l_a$ is the minimum
among all the $\{\l_a\}$: or in other words $\min_a \l_a = \l_a  \d_{a,{\rm min}}$.
Then we have
\beq\label{Fspexpl}
\begin{split}
\mathcal F[\hat\y^{\rm 1RSB}] 
&=\lim_{n\to 0}\sum_{n_1,\ldots, n_m; \sum_a^mn_a=n}\frac{n!}{n_1!\ldots n_m!}\exp\left[ -\frac 12 \sum_{a=1}^m \frac{n_a}{n}\y_{aa}^{\rm 1RSB}+\frac{1}{2} \sum_{a,b}^{1,m}\frac{n_an_b}{n^2}\y_{ab}^{\rm 1RSB}  \right]\\
&=\lim_{n\to 0}e^{-\wh A}\sum_{n_1,\ldots, n_m; \sum_a^mn_a=n}\frac{n!}{n_1!\ldots n_m!}\exp\left[ \wh A \sum_{a=1}^m \frac{n_a^2}{n^2} \right]\\
&=\lim_{n\to 0}e^{-\wh A}\sum_{n_1,\ldots, n_m; \sum_a^mn_a=n}\frac{n!}{n_1!\ldots n_m!}\int \left(\prod_{a=1}^m\frac{\de \lambda_a}{\sqrt{2\pi}}\right)\exp\left[ - \sum_{a=1}^m \frac{\lambda_a^2}{2} -\sqrt{2 \wh A}\sum_{a=1}^m\frac{n_a\lambda_a}{n}\right]\\
&=\int \left(\prod_{a=1}^m\frac{\de \lambda_a}{\sqrt{2\pi}}\right)\exp\left[ - \sum_{a=1}^m \frac{\lambda_a^2}{2} -\sqrt{2 \wh A}\min_{a}\lambda_a - \wh A\right]\\
&=\int \left(\prod_{a=1}^m\frac{\de \lambda_a}{\sqrt{2\pi}}\right)\exp\left[ - \frac12 \sum_{a=1}^m \left(\l_a + \sqrt{2 \wh A} \, \d_{a,{\rm min}} \right)^2 \right]\\
&=m\int_{-\infty}^\infty \frac{\de \lambda_1}{\sqrt{2\pi}} e^{-\frac 12 \left(\l_1 + \sqrt{2 \wh A} \right)^2}\left[ \int_{\lambda_1}^\infty \frac{\de \lambda}{\sqrt {2 \pi}} e^{-\frac12\lambda^2}    \right]^{m-1}\\
&=m\int_{-\infty}^\infty \frac{\de \lambda_1}{\sqrt{2\pi}} e^{-\frac 12 \left(\l_1 + \sqrt{2 \wh A} \right)^2}
\left[\Th\left( -\frac{\l_1}{\sqrt{2}} \right) \right]^{m-1}
\end{split}
\eeq
where
\beq
\mathrm{erf}(x)=\frac{2}{\sqrt \pi}\int_0^x\de y\, e^{-y^2}\ \ \ \ \ \ \ \ \ \ \ \ 
\Theta(x)=\frac 12 +\frac 12 \mathrm{erf}(x) = \frac{1}{\sqrt \pi}\int_{-x}^\io \de y\, e^{-y^2}
\eeq
We also define for later convenience
\beq\label{Thk}
\Th_k(x) = \frac{1}{\sqrt {2 \pi}}\int_{x}^\io \de y\, y^k \, e^{-\frac12 y^2}
\eeq
Note that 
\beq\begin{split}
\Th_0(x) &= \Th(-x/\sqrt{2}) \\
\Th_1(x) &= e^{-\frac12 x^2}/\sqrt{2 \pi} \\
\Th_2(x) &= x \Th_1(x) + \Th(-x/\sqrt{2}) \\
\Th_3(x) &= \Th_1(x) (2 + x^2) \\
\Th_4(x) &= x (3+x^2) \Th_1(x) + 3 \Th_0(x) \\
\end{split}\eeq
and so on. Eq.~\eqref{Fspexpl} provides the derivation of the interaction part of Eq.~\eqref{pap1:1RSB}~\cite{PZ10,KPZ12}.

\subsubsection{Monomials of $n$}

We now want to expand the quantity $\mathcal F$ around the 1RSB solution. 
The part of the Hessian
matrix coming from the interaction term is: 
\begin{gather}
M^{(I)}_{a<b;c<d} = \frac{\delta^2 \mathcal F}{\delta \y_{a<b}\delta \y_{c<d}}[\hat v^{\rm 1RSB}]=
\lim_{n\to 0}\sum_{n_1,\ldots, n_m; \sum_a^mn_a=n}\frac{n!}{n_1!\ldots n_m!}\,  f(n_a,n_b) f(n_c,n_d)  \exp\left[  -\wh A +\wh A\sum_{a=1}^m \frac{n_a^2}{n^2}  \right] 
\end{gather}
where the functions $f$ are at most quadratic:
\begin{gather}
f(n_a,n_b)=\frac{n_a}{2n}+\frac{n_b}{2n}- \frac{n_a^2}{2n^2} -\frac{n_b^2}{2n^2}+\frac{n_an_b}{n^2}\:.
\end{gather}
By introducing the following notation
\beq
\langle O  \rangle=\lim_{n\to 0}\sum_{n_1,\ldots, n_m; \sum_a^mn_a=n}\frac{n!}{n_1!\ldots n_m!}\, O \exp\left[  -\wh A +\wh A\sum_{a=1}^m \frac{n_a^2}{n^2}  \right] 
\eeq
we have that the Hessian matrix is given by
\beq
\begin{split}
M^{(I)}_{a<b;c<d}&=\frac{1}{4}\left(\langle \frac{n_a n_c}{n^2}\rangle +\langle \frac{n_a n_d}{n^2}\rangle+\langle \frac{n_b n_c}{n^2}\rangle +\langle \frac{n_b n_d}{n^2}\rangle\right)\\
&-\frac{1}{4}\left(\langle \frac{n_a^2 n_c}{n^3}\rangle +\langle \frac{n_a^2 n_d}{n^3}\rangle+\langle \frac{n_b^2 n_c}{n^3}\rangle +\langle \frac{n_b^2 n_d}{n^3}\rangle+\langle \frac{n_a n_c^2}{n^3}\rangle +\langle \frac{n_a n_d^2}{n^3}\rangle+\langle \frac{n_b n_c^2}{n^3}\rangle +\langle \frac{n_b n_d^2}{n^3}\rangle \right.\\
&\left.-2\langle \frac{n_an_cn_d}{n^3}\rangle-2\langle \frac{n_bn_cn_d}{n^3}\rangle-2\langle \frac{n_cn_an_b}{n^3}\rangle-2\langle \frac{n_dn_an_b}{n^3}\rangle\right)\\
&+\frac{1}{4}\left( \langle \frac{n_a^2 n_c^2}{n^4}\rangle +\langle \frac{n_a^2 n_d^2}{n^4}\rangle+\langle \frac{n_b^2 n_c^2}{n^4}\rangle +\langle \frac{n_b^2 n_d^2}{n^4}\rangle-2\langle\frac{n_a^2n_cn_d}{n^4}\rangle-2\langle\frac{n_b^2n_cn_d}{n^4}\rangle -2\langle\frac{n_c^2n_an_b}{n^4}\rangle\right.\\
&\left.-2\langle\frac{n_d^2n_an_b}{n^4}\rangle+4\langle \frac{n_an_b n_cn_d}{n^4}\rangle  \right) \ .
\end{split}
\eeq
Hence we want to compute averages of monomials of the $\{n_a\}$, which can be written as follows:
\beq\label{ntolambda}
\begin{split}
&\left\langle \frac{n_{a_1}}n \cdots  \frac{n_{a_k}}n \right\rangle
=\lim_{n\to 0}\frac{1}{n^k}\sum_{n_1,\ldots, n_m; \sum_a^mn_a=n}\frac{n!}{n_1!\ldots n_m!} \,
n_{a_1} \cdots n_{a_k}
\exp\left[ -\wh A + \wh A \sum_{a=1}^m \frac{n_a^2}{n^2}   \right]\\
&=e^{-\wh A}\lim_{n\to 0}\int \left(\prod_a \frac{\de \lambda_a}{\sqrt{2\pi}}\right)
e^{-\sum_{a=1}^m\frac{\lambda_a^2}{2}}\sum_{n_1,\ldots, n_m; \sum_a^mn_a=n}\frac{n!}{n_1!\ldots n_m!}
\frac{n_{a_1} \cdots n_{a_k}}{n^k}
\exp\left[ -\sqrt{2 \wh A} \sum_{a=1}^m \frac{n_a\lambda_a}{n}   \right]\\
&=\frac{ (-1)^k e^{-\wh A}}{(2 \wh A)^{k/2}}
\lim_{n\to 0}\int \left(\prod_a \frac{\de \lambda_a}{\sqrt{2\pi}}\right)
e^{-\sum_{a=1}^m\frac{\lambda_a^2}{2}}
\sum_{n_1,\ldots, n_m; \sum_a^mn_a=n}\frac{n!}{n_1!\ldots n_m!}
\frac{\partial^k}{\partial \lambda_{a_1} \cdots \partial \lambda_{a_k}}
\exp\left[ -\sqrt{2 \wh A} \sum_{a=1}^m \frac{n_a\lambda_a}{n}   \right]\\
&=\frac{ e^{-\wh A}}{(2 \wh A)^{k/2}}\int \left(\prod_{a} \frac{\de \lambda_a}{\sqrt{2\pi}}\right)\left(\frac{\partial^k}{\partial \lambda_{a_1} \cdots \partial \lambda_{a_k}}e^{-\sum_{a=1}^m\frac{\lambda_a^2}{2}}\right)\exp\left[  -\sqrt{2 \wh A} \min_a \lambda_a  \right]\\
&=\frac{1}{(2\wh A)^{k/2}}\int \left(\prod_{a} \frac{\de \lambda_a}{\sqrt{2\pi}}\right)
\left(e^{\sum_{a=1}^m\frac{\lambda_a^2}{2}}\frac{\partial^k}{\partial \lambda_{a_1} \cdots \partial \lambda_{a_k}}e^{-\sum_{a=1}^m\frac{\lambda_a^2}{2}}\right)
\exp\left[ - \frac12 \sum_{a=1}^m \left(\l_a + \sqrt{2 \wh A} \, \d_{a,{\rm min}} \right)^2 \right]\\
&=\frac{1}{(2\wh A)^{k/2}} \left\langle
e^{\sum_{a=1}^m\frac{\lambda_a^2}{2}}\frac{\partial^k}{\partial \lambda_{a_1} \cdots \partial \lambda_{a_k}}e^{-\sum_{a=1}^m\frac{\lambda_a^2}{2}} \right\rangle
\end{split}
\eeq
where the definition of the average has been changed to
\beq
\la O \ra = \int \left(\prod_{a} \frac{\de \lambda_a}{\sqrt{2\pi}}\right)
O
\exp\left[ - \frac12 \sum_{a=1}^m \left(\l_a + \sqrt{2 \wh A} \, \d_{a,{\rm min}} \right)^2 \right]
\eeq
The factor in parenthesis is a polynomial in $\{\l_a\}$, hence we now want to be able to write averages of
monomials of $\l$. Using the replica symmetry of the average over $\{\l_a\}$,
we need in particular the following objects:
\beq\label{BTD}
\begin{split}
B_{ab} &= \la \frac{n_a n_b}{n^2} \ra = \frac{1}{2\wh A} \la -\d_{ab} + \l_a\l_b  \ra \\
T_{abc} &=  \la \frac{n_a n_b n_c}{n^3} \ra =\frac{1}{(2\wh A)^{3/2}} 
\la \d_{bc} \l_a + \d_{ac} \l_b + \d_{ab} \l_c - \l_a \l_b \l_c \ra \\
\D_{abcd} &= \la \frac{n_a n_b n_c n_d}{n^4} \ra =
\frac{1}{(2\wh A)^{2}} 
\langle 
\d_{ab} \d_{cd} + \d_{bc} \d_{ad} + \d_{ac} \d_{cd}
- \d_{ab} \l_c\l_d - \d_{bc} \l_a\l_d - \d_{ac} \l_b\l_d \\
&\hskip100pt - \d_{cd} \l_a\l_b - \d_{bd} \l_a\l_c - \d_{ad} \l_b\l_c
+ \l_a \l_b \l_c \l_d
\rangle \\
\end{split}\eeq
and the Hessian matrix is
\beq
\begin{split}
M^{(I)}_{a<b;c<d}&=\frac{1}{4}\left(
B_{ac} + B_{ad} + B_{bc} + B_{bd}
\right)\\
&-\frac{1}{4}\left(
T_{aac} + T_{aad} + T_{bbc} + T_{bbd} + T_{acc} + T_{add} + T_{bcc} + T_{bdd}
-2 T_{acd} - 2 T_{bcd} -2 T_{abc} - 2 T_{abd} 
\right)\\
&+\frac{1}{4}\left(
\D_{aacc} + \D_{aadd} + \D_{bbcc}
+ \D_{bbdd} 
-2 \D_{aacd} -2 \D_{bbcd} -2 \D_{ccab}  -2 \D_{abdd}
+4 \D_{abcd}
\right)
\end{split}
\eeq

\subsubsection{Monomials of $\l$}

We therefore need to compute several monomials of the $\{\l_a\}$, which are listed in the following.
Calculations follow Eq.~\eqref{Fspexpl} and it will be convienient to define one more average over $\l$
as
\beq
\la f(\l) \ra = \int_{-\infty}^\infty \frac{\de \lambda}{\sqrt{2\pi}} 
e^{-\frac 12 \left(\lambda+\sqrt{2 \wh A}\right)^2} f(\l)
\eeq
Then we have
\beq\label{monolambda}
\begin{split}
\la 1 \ra =& \FF[\hat v^{\rm 1RSB}] = \la m \, \Th_0(\l)^{m-1} \ra \\
\la \l_1^k \ra =& \la \l^k \, \Th_0(\l)^{m-1} + (m-1) \, \Th_k(\l) \Th_0(\l)^{m-2} \ra \\
\la \l_1^k \l_2^l \ra =& \la
\l^k \Th_l(\l) \Th_0(\l)^{m-2} 
+ \l^l  \Th_k(\l) \Th_0(\l)^{m-2}
+ (m-2) \Th_k(\l) \Th_l(\l) \Th_0(\l)^{m-3} \ra \\
\la \l_1^k \l_2^l \l_3^n \ra =& \langle
\l^k \Th_l(\l) \Th_n(\l) \Th_0(\l)^{m-3} +
\l^l \Th_k(\l) \Th_n(\l) \Th_0(\l)^{m-3} +
\l^n \Th_k(\l) \Th_l(\l) \Th_0(\l)^{m-3} \\ &+
(m-3) \Th_k(\l) \Th_l(\l)  \Th_n(\l) \Th_0(\l)^{m-4}
\rangle \\
\la \l_1^k \l_2^l \l_3^n \l_4^p \ra =& \langle
\l^k \Th_l(\l) \Th_n(\l) \Th_p(\l) \Th_0(\l)^{m-4} +
\l^l \Th_k(\l) \Th_n(\l) \Th_p(\l) \Th_0(\l)^{m-4} \\
&+ \l^n \Th_k(\l) \Th_l(\l) \Th_p(\l) \Th_0(\l)^{m-4} +
\l^p \Th_k(\l) \Th_l(\l) \Th_n(\l) \Th_0(\l)^{m-4} 
\\ &+
(m-4) \Th_k(\l) \Th_l(\l)  \Th_n(\l) \Th_p(\l) \Th_0(\l)^{m-5}
\rangle \\
\end{split}
\eeq
and so on.

\subsubsection{The structure of the mass matrix}

Thanks to replica symmetry the Hessian matrix has only three independent matrix elements.
These are
\beq\begin{split}
M^{(I)}_{12;12} &= \frac12 ( B_{11} + B_{12} ) + (T_{112} - T_{111} ) + 
\frac14 ( 2 \D_{1111} + 6 \D_{1122} - 8 \D_{1112} )   \ , \\
M^{(I)}_{12;13} &= \frac14 ( B_{11} + 3 B_{12})  
- \frac12 (  T_{111} +  T_{112} - 2 T_{123} ) 
+ \frac14 ( \D_{1111} + 3 \D_{1122} -4 \D_{1112}  ) \ , \\
M^{(I)}_{12;34} &= B_{12} - 2 ( T_{112} - T_{123}) + 
(\D_{1122} - 2 \D_{1123} +  \D_{1234} ) \ .
\end{split}\eeq
It is however convenient to write the matrix in this form:
\beq
\begin{split}
M_{a<b;c<d}^{(I)}&=
M_1^{(I)}\frac{\delta_{ac}\delta_{bd}+\delta_{ad}\delta_{bc}}{2}+
M_2^{(I)}\frac{\delta_{ac}+\delta_{ad}+\delta_{bc}+\delta_{bd}}{4}+M_3^{(I)}
\end{split}
\eeq
where
\beq
\begin{array}{lll}
M^{(I)}_1 & = 2 M^{(I)}_{12;12} - 4 M^{(I)}_{12;13} +2 M^{(I)}_{12;34} &= 
2 \D_{1122} - 4 \D_{1123}+ 2\D_{1234} \\
M^{(I)}_2 & = 4 M^{(I)}_{12;13} - 4 M^{(I)}_{12;34} &= 
B_{11}-B_{12}-2 T_{111}+6 T_{112}-4 T_{123} \\
& & \hskip10pt +\D_{1111} - 4 \D_{1112} - \D_{1122} + 8 \D_{1123} - 4 \D_{1234}
\\
M^{(I)}_3 & = M^{(I)}_{12;34} &=    B_{12} - 2  T_{112} +2 T_{123}) + 
\D_{1122} - 2 \D_{1123} +  \D_{1234}      \\ 
\end{array}
\eeq
The above equation, together with Eqs.~\eqref{BTD} and \eqref{monolambda}, 
give the complete expression of the interaction part of the Hessian matrix.

\subsection{The replicon}

\begin{figure}
\includegraphics[width=.5\textwidth]{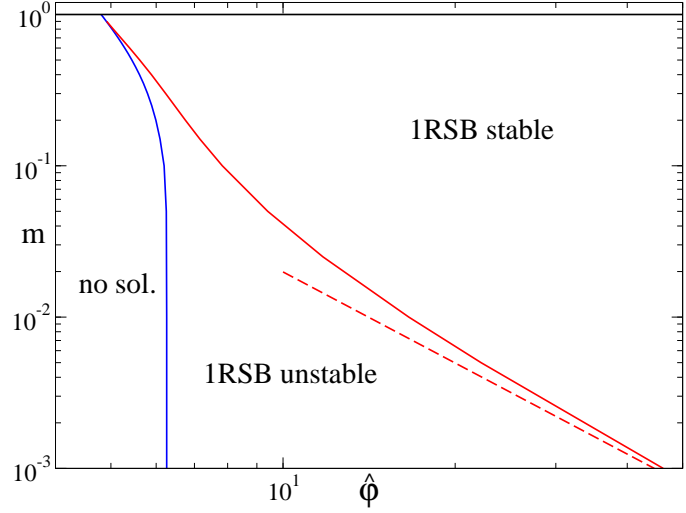}
\caption{
Phase diagram of the 1RSB solution in the $(m,\wh\f)$ plane (to compare with Fig.~\ref{pspi}, recall that rougly speaking
pressure is the inverse of $m$), including its instability.
The black line at $m=1$ corresponds to the liquid phase.
The blue line is the dynamical line $\wh\f_d(m)$ at which a non-trivial 1RSB solution appears.
The red line is the instability line $\wh\f_G(m)$ where the replicon vanishes. The red dashed
line is its asymptotic behavior for small $m$ and large $\wh\f$.
}
\label{fig:mf}
\end{figure}

The stability of the 1RSB solution depends crucially on the replicon eigenvalue,
$\l_R = M_1 =\wh A^2 M_1^{(E)} - 4 \wh \f \, \wh A^2 M^{(I)}_1$.
Collecting all the above results and simplifying some terms we get
\beq\begin{split}
\l_R &= -4 -2 \wh \f \, \L_m(\wh A) \\
\L_m(\wh A) &=
\la
\Th_0(\l)^{m-5} [\Th_1(\l)^2 - \l \Th_1(\l) \Th_0(\l)]
[(2 - 2 \l^2) \Th_0( \l)^2 + (m-4) \Th_1(\l)^2 + 
(6 - m) \l  \Th_0(\l) \Th_1(\l)]
\ra \\
&=
\la
\Th_0(\l)^{m-1} \left[\left(\frac{\Th_1(\l)}{\Th_0(\l)}\right)^2 - \l \frac{\Th_1(\l)}{\Th_0(\l)}\right]
\left[(2 - 2 \l^2) + (m-4) \left(\frac{\Th_1(\l)}{\Th_0(\l)}\right)^2 + 
(6 - m) \l \frac{\Th_1(\l)}{\Th_0(\l)}\right]
\ra 
\end{split}\eeq
We can compute this numerically on the 1RSB solution, where $\wh A$ is the solution of
Eq.~\eqref{pap1:1RSBeq}, 
to find the point where
$\l_R=0$ and the 1RSB solution becomes unstable.
The instability curve in the $(m,\wh\f)$ 
plane, which corresponds to the line $\wh\f_G(m)$ on which the replicon vanishes, 
is reported in Fig.~\ref{fig:mf}.

Asymptotically we obtain $\wh\f_{G}(m) \sim m^{-1/2}$ for $m\to 0$.
To explain this we must investigate the asymptotics of the function
$\L(m,\wh A)$ when both $m$ and $\wh A$ are small.
It is convenient to use Eq.~\eqref{pap1:1RSBeq} to eliminate $\wh\f$ instead of $\wh A$. Doing
this the equation for the instability becomes
\beq\label{FmLm}
2 \FF_m(\wh A) = - \L_m(\wh A) \ ,
\eeq
which must be solved to obtain $\wh A_G(m)$ and then $\wh \f_G(m)$ using Eq.~\eqref{pap1:1RSBeq}.
We want to show that $\wh A_G(m) \sim m^2$, and that this implies $\wh\f_{G}(m) \sim m^{-1/2}$.

First of all let us examine the asymptotics of the different terms
for large and positive $\l$. For $\l\to\io$ we have
\beq\label{Las1}
\begin{split}
\Th_0(\l) &\sim \frac{e^{-\frac12 \l^2}}{\sqrt{2\pi} \l} \left( 1 - \frac1{\l^2} + \frac{3}{\l^4} - \frac{15}{\l^6} + \cdots \right) \ , \\
\left(\frac{\Th_1(\l)}{\Th_0(\l)}\right)^2 - \l \frac{\Th_1(\l)}{\Th_0(\l)} &\sim 1 - \frac1{\l^2} + \frac{6}{\l^4} - \frac{50}{\l^6} + \cdots \ , \\
(2 - 2 \l^2) + (m-4) \left(\frac{\Th_1(\l)}{\Th_0(\l)}\right)^2 + 
(6 - m) \l \frac{\Th_1(\l)}{\Th_0(\l)} &\sim
m - \frac{m}{\l^2} + \frac{6m-4}{\l^4} + \frac{52 - 50 m}{\l^6} + \cdots
\ .
\end{split}\eeq
It will be convenient for the following to define
\beq\begin{split}
\LL(\l) &= 
\left[\left(\frac{\Th_1(\l)}{\Th_0(\l)}\right)^2 - \l \frac{\Th_1(\l)}{\Th_0(\l)}\right]
\left[2 - 2 \l^2 + (m-4) \left(\frac{\Th_1(\l)}{\Th_0(\l)}\right)^2 + 
(6 - m) \l \frac{\Th_1(\l)}{\Th_0(\l)}\right] \ , \\
\LL(\l) &\sim m - \frac{2m}{\l^2} + \frac{13m-4}{\l^4} +\frac{56-112 m}{\l^6} + \cdots \ .
\end{split}\eeq

Now we expand Eq.~\eqref{FmLm} at small $\wh A$. 
Let us recall that $\GG_m(\wh A) = 1 - m \la \Th_0(\l)^{m-1} \ra$.
From this we obtain
\beq\begin{split}
\GG_m(\wh A) &= G_1(m) \sqrt{\wh A} + G_2(m) \wh A + \cdots \ , \\
\FF_m(\wh A) &= \frac{G_1(m)}{1-m} \frac12 \sqrt{\wh A} + \frac{G_2(m)}{1-m} \wh A + \cdots \ , \\
G_1(m) &= -m \int \frac{\de\l}{\sqrt{2\pi}} e^{-\l^2/2} \Th_0(\l)^{m-1} (-\l \sqrt{2}) \ , \\
G_2(m) &= -m \int \frac{\de\l}{\sqrt{2\pi}} e^{-\l^2/2} \Th_0(\l)^{m-1} ( \l^2 - 1) \ ,
\end{split}\eeq
and similarly (the fact that the horrible integral corresponding to $\L_m(\wh A=0)$ is exactly 0 
can be proven by a series of integrations by parts):
\beq\begin{split}
\L_m(\wh A) &= L_1(m) \sqrt{\wh A} + L_2(m) \wh A + \cdots \ , \\
L_1(m) &=  \int \frac{\de\l}{\sqrt{2\pi}} e^{-\l^2/2} \Th_0(\l)^{m-1} \LL(\l)  (-\l \sqrt{2}) \ , \\
L_2(m) &=  \int \frac{\de\l}{\sqrt{2\pi}} e^{-\l^2/2} \Th_0(\l)^{m-1} \LL(\l) ( \l^2 - 1) \ .
\end{split}\eeq
Hence Eq.~\eqref{FmLm} becomes
\beq
0 = \sqrt{\wh A} \D_1(m) + \wh A \D_2(m) + \cdots
\hskip10pt
\Rightarrow
\hskip10pt
\sqrt{\wh A_G} = - \frac{\D_1(m)}{\D_2(m)} + \cdots \ ,
\eeq
with
\beq\begin{split}
\D_1(m) &=  \frac{G_1(m)}{1-m} + L_1(m) \ , \\
\D_2(m) &= 2 \frac{G_2(m)}{1-m} + L_2(m) \ . 
\end{split}\eeq
Asymptotically for small $m$, the integrals in the above expressions can have different behaviors,
depending on the behavior of the integrand for large $\l$ when $m\to 0$. In fact, if the integrand decays
faster than $1/\l$, the integral is well defined and has a finite limit for $m\to 0$. In the opposite case,
the integral is divergent and the divergence is dominated by the large $\l$ behavior: in this case one has to analyze the
possibly divergent part to determine the behavior of the integral at $m\to 0$.
In the case of $\D_1(m)$, thanks to a subtle cancellation, the large $\l$ contribution to the integral is 
\beq
\D_1(m) \sim  \int \frac{\de\l}{\sqrt{2\pi}} e^{- m\l^2/2} (\sqrt{2\pi} \l)^{1-m} (-\l \sqrt{2}) \left[ - m^2 - \frac{2m}{\l^2} - \frac{4}{\l^4} + \cdots \right] \ .
\eeq
The first two terms give contributions that are not divergent when $m\to 0$, hence they are subleading with respect to the $1/\l^4$ term that
gives a finite contribution.
We conclude that $\D_1(m)$ has a finite limit given by
\beq
\D_1(0) = \int \frac{\de\l}{\sqrt{2\pi}} e^{-\l^2/2} \Th_0(\l)^{-1} \left[\left(\frac{\Th_1(\l)}{\Th_0(\l)}\right)^2 - \l \frac{\Th_1(\l)}{\Th_0(\l)}\right]
\left[2 - 2 \l^2 -4 \left(\frac{\Th_1(\l)}{\Th_0(\l)}\right)^2 + 
6  \l \frac{\Th_1(\l)}{\Th_0(\l)}\right]  (-\l \sqrt{2}) \approx 1.6
\eeq
Instead, the leading large $\l$ behavior of the integrand of $\D_2(m)$ is, changing variable to $y = \sqrt{m} \l$:
\beq\begin{split}
\D_2(m) &\sim  \int \frac{\de\l}{\sqrt{2\pi}} e^{- m\l^2/2} (\sqrt{2\pi} \l)^{1-m} (\l^2-1) \left[ - m + \cdots \right] \\
& = -\frac1{m} \int_0^\io \de y \, e^{-y^2/2} \, y^3 = -\frac2m \ .
\end{split}\eeq
We conclude that $\sqrt{\wh A_G} \approx 0.8 m$. Finally, we can show similarly that for small $m$
\beq\begin{split}
G_1(m) &\sim -m \int \frac{\de\l}{\sqrt{2\pi}} e^{-m\l^2/2} (\sqrt{2\pi} \l)^{1-m} (-\l \sqrt{2}) \\
& =  \sqrt{\frac2m} \int_0^\io \de y \, e^{-y^2/2} y^2 \sim \sqrt{\frac{\pi}{m}} \ , 
\end{split}\eeq
hence
\beq
\wh \f_G \sim \frac{1}{\frac{G_1(m)}{1-m} \frac12 \sqrt{\wh A_G} } \sim
\sqrt{ \frac{4m}{\pi \wh A_G}} \approx \sqrt{ \frac{4}{\pi }} \frac{1}{0.8 \sqrt{m}} \approx 1.41 m^{-1/2} \ . 
\eeq
Both asymptotic results for $\wh A_G$ and $\wh\f_G$ are perfectly consistent with the numerical data.

\subsection{The 2RSB solution}

When the 1RSB solution becomes unstable, one must consider further RSB.
We performed a 2RSB calculation. 
Then we can linearize the 2RSB solution close to the 1RSB one and obtain the line at which
the 2RSB provides a better maximization of the entropy, hence becoming stable.
This provides an independent calculation of the 1RSB instability, which we verified to be coherent
with the one reported above.
A complete characterization of the 2RSB solution (as well as the 3, 4, $\cdots$, $\io$RSB ones) will be
presented in future papers of this series.

\section{Computation of the dynamic exponents from the cubic expansion}
\label{app:MCT}

The same strategy allows to obtain the cubic terms in the expansion. From these, following the procedure of Ref.~\cite{CFLPRR12,FJPUZ13}, 
one can compute the mean-field dynamical critical exponents at the dynamical glass transition, the so-called exponent parameter of
Mode-Coupling theory, $\l_{\rm MCT}$.
Although this calculation is not the main scope of this paper, we report it in this section.

Let us define, following the same notation as for the second order terms (hence for $a\neq b$, $c\neq d$, $e\neq f$ which we omit from now on)
\beq
W_{ab,cd,ef} = \frac{\d^3 s[\hat \a]}{\d \a_{a<b} \d\a_{c<d} \d\a_{e<f}} \ .
\eeq
Exploiting the replica symmetry, the two coefficients $w_1$ and $w_2$ can be written in the following form
\beq\label{DeDominicis}
\begin{split}
w_1&=W_{ab,bc,ca}-3W_{ab,ac,bd}+3W_{ac,bc,de}-W_{ab,cd,ef} \ , \\
w_2&=\frac{1}{2}W_{ab,ab,ab}-3W_{ab,ab,ac}+\frac{3}{2}W_{ab,ab,cd}+3W_{ab,ac,bd}+2W_{ab,ac,ad}-6W_{ac,bc,de}+2W_{ab,cd,ef} \ ,
\end{split}
\eeq
and we then have
$\l_{\rm MCT} = \frac{w_2}{w_1}$.

Defining
\beq\begin{split}
W^{(I)}_{ab,cd,ef} & = \left. \frac{\delta^3 \mathcal F[\hat\y]}{\delta \y_{a<b}\delta \y_{c<d} \d\y_{e<f}} 
\right|_{\hat \y = 2 \hat \a^{\rm 1RSB}} \\
W_{ab,cd,ef}^{(E)}&=\left.\frac{\delta^3}{\delta \a_{a<b}\delta\a_{c<d}\d\a_{e<f}}\log\det( \hat\alpha^{m,m} ) 
\right|_{\hat \a = \hat \a^{\rm 1RSB}}
\end{split}\eeq
we have
\beq
W_{ab,cd,ef} =W_{ab,cd,ef}^{(E)}-8 \wh \varphi\, W_{ab,cd,ef}^{(I)} \ ,
\eeq
hence we have a similar relation for $w_1$ and $w_2$.
We now compute these two terms separately.

\subsection{The entropic term}

Following the same strategy as in Sec.~\ref{Mentropic}, we obtain
\beq
W^{(E)}_{ab,cd,ef} = 
\Tr \left[ 
\hat\b^{-1}  \frac{\delta \hat\b}{\delta \a_{a<b}}  \hat\b^{-1} \frac{\delta \hat\b}{\delta \a_{c<d}} \hat\b^{-1} \frac{\delta \hat\b}{\delta \a_{e<f}}
\right]_{\hat \b = \hat \b^{\rm 1RSB}}
+
\Tr \left[ 
\hat\b^{-1}  \frac{\delta \hat\b}{\delta \a_{a<b}}  \hat\b^{-1} \frac{\delta \hat\b}{\delta \a_{e<f}} \hat\b^{-1} \frac{\delta \hat\b}{\delta \a_{c<d}}
\right]_{\hat \b = \hat \b^{\rm 1RSB}}
\eeq
Using Eq.~\eqref{DeDominicis}, the results of Sec.~\ref{Mentropic} and performing the traces we obtain
\beq\begin{split}
w_1^{(E)} &= \frac2{\wh A^3} \ , \\
w_2^{(E)} &= 0 \ .
\end{split}\eeq

\subsection{The interaction term}

The interaction term is
\beq
W^{(I)}_{ab,cd,ef} =\left. \frac{\d^3 \FF[\hat \y]}{\d \y_{a<b} \d\y_{c<d} \d\y_{e<f}} \right|_{\hat\y = 2\hat\a^{\rm 1RSB}} = \la f(n_a,n_b) f(n_c,n_d) f(n_e,n_f) \ra \ .
\eeq
Using the expression of $f$, expanding the products, and simplifying many monomials using the symmetries
(e.g. $\la n_a^3 n_d n_f^2 \ra = \la n_a^3 n_b^2 n_c \ra$), 
we obtain
\beq\begin{split}
w_1^{(I)} &= \la \frac{n_a^2 n_b^2 n_c^2 - 3 n_a^2 n_b^2 n_c n_d + 3 n_a^2 n_b n_c n_d n_e - 
 n_a n_b n_c n_d n_e n_f}{n^6} \ra \ , \\
w_2^{(I)} &= \la \frac{ 
(1/2) n_a^3 n_b^3 - 3 n_a^3 n_b^2 n_c + 2 n_a^3 n_b n_c n_d + 
 (9/2) n_a^2 n_b^2 n_c n_d - 6 n_a^2 n_b n_c n_d n_e + 2 n_a n_b n_c n_d n_e n_f
 }{n^6} \ra \ .
\end{split}\eeq
Now we convert the average over $n$ into an average over $\l$, using Eq.~\eqref{ntolambda}.
Performing the derivatives and exploiting similar symmetries to simplify the result we obtain
\beq\begin{split}
w_1^{(I)} &= 
\frac1{(2 \wh A)^3} \la -1 + 3 \l_1^2 - 3 \l_1 \l_2 - 
 3 \l_1^2 \l_2^2 + 
 6 \l_1^2 \l_2 \l_3 + \l_1^2 \l_2^2 
\l_3^2 - 3 \l_1 \l_2 \l_3 \l_4 - 
 3 \l_1^2 \l_2^2 \l_3 \l_4 + 
 3 \l_1^2 \l_2 \l_3 \l_4 \l_5 -
\l_1 \l_2 \l_3 \l_4 \l_5 \l_6
\ra
 \ , \\
w_2^{(I)} &= 
\frac1{(2 \wh A)^3} \la
\frac12 \l_1^3 \l_2^3 - 
 3 \l_1^3 \l_2^2 \l_3 + 
 2 \l_1^3 \l_2 \l_3 \l_4 + 
 \frac92 \l_1^2 \l_2^2 \l_3 \l_4 - 
 6 \l_1^2 \l_2 \l_3 \l_4 \l_5 + 
 2 \l_1 \l_2 \l_3 \l_4 \l_5 \l_6
\ra
 \ .
\end{split}\eeq
This result, together with Eq.~\eqref{monolambda}, allows for the explicit computation of these terms.
After some simplifications, we obtain
\beq\begin{split}
w_1^{(I)} = 
-\frac1{(2 \wh A)^3} \Big\langle &
\Theta_0(\l)^{m-7} [\Theta_0(\l)^2 + \Theta_1(\l)^2 - \Theta_0(\l) \Theta_2(\l)]^2 \\
& \{(m -  3 \l^2) \Theta_0(\l)^2 + (m-6 ) \Theta_1(\l)^2 + \Theta_0(\l) [6 \l \Theta_1(\l) - (m -3) \Theta_2(\l)]\}
\Big\rangle
 \ , \\
w_2^{(I)} = 
\frac1{2(2 \wh A)^3} \Big\langle &
 \Theta_0(\l)^{m-7}  [2 \Theta_1(\l)^3 - 
   3 \Theta_0(\l) \Theta_1(\l) \Theta_2(\l) + \
\Theta_0(\l)^2 \Theta_3(\l)] \\ 
&\{2 \l^3 \
\Theta_0(\l)^3 + 2 (m-6) \Theta_1(\l)^3 + 
   3 \Theta_0(\l) \Theta_1(\l) (4 \l 
\Theta_1(\l) - (m-4) \Theta_2(\l))  \\ &+ \Theta_0(\l)^2 [-6 
\l (\l \Theta_1(\l) + \Theta_2(\l)) + 
(m-2) \Theta_3(\l)]\}
\Big\rangle
 \ ,
\end{split}\eeq
which can be easily computed numerically.

\subsection{Numerical result}

Collecting all the terms together we obtain the final result
\beq
\l_{\rm MCT} = \frac{w_2}{w_1} = \frac{-8 \wh \f w_2^{(I)}}{2/\wh A^3 - 8 \wh \f w_1^{(I)} } \ .
\eeq
When computed at the dynamical transition with $m=1$, $\wh\f = \wh\f_d = 4.80677$ and $\wh A = \wh A_d = 0.57668$ given by
the solution of Eq.~\eqref{pap1:1RSBeq}, we obtain
\beq
\l_{\rm MCT} =0.70698 \ ,
\eeq
which implies that the MCT exponents are $a=0.324016$, $b=0.629148$ and $\g = 2.33786$.
The result for $\g$ is roughly consistent with the numerical estimate of Ref.~\cite{CIPZ12}.

\section{Phenomenological extension to finite dimensions}
\label{sec:IV}

We can obtain quantitative results in finite $d$ by a phenomenological extension of Eq.~\eqref{eq:gauss_r}.
First we go back to non-rescaled density and we rearrange it as
\beq
s[\hat \a]  = 1 - \log\r -  2^{d-1} \f
+ d \log m + \frac{(m-1)d}{2} \log(2 \pi e D^2/d^2) + \frac{d}2 \log \det(\hat \a^{m,m}) + 2^{d-1} \f \,
[1- \FF\left( 2 \hat \a \right)] \ .
\eeq
We now recognize that $s_{liq} = 1 - \log\r -  2^{d-1} \f$. Furthermore, by comparison with the finite $d$ results obtained
in the small cage expansion~\cite{PZ10}, we know that the interaction term is renormalized
by the contact value of the liquid correlation $y_{liq}(\f)$. We therefore can propose the following form for the entropy:
\beq
\label{eq:gauss_r_finited}
s[\hat \a]  = s_{liq}(\f)
+ d \log m + \frac{(m-1)d}{2} \log(2 \pi e D^2/d^2) + \frac{d}2 \log \det(\hat \a^{m,m}) + 2^{d-1} \f y_{liq}(\f) \,
[1- \FF\left( 2 \hat \a \right)] \ .
\eeq
In the 1RSB scheme we obtain
\beq
\begin{split}
&s[\wh A] = 
s_{liq}(\f) + \frac{d}{2} \log m + \frac{(m-1)d}{2} + \frac{(m-1)d}{2} \log\left(\frac{2 \pi D^2 \wh A}{d^2}\right) 
+ 2^{d-1} \f y_{liq}(\f) \, \GG_m(\wh A)  \ , \\
&\frac{d}{2^d \f y_{liq}(\f)} = \FF_m(\wh A) \ , \\
&\l_R = -4 -2 \frac{2^d \f y_{liq}(\f)}d \, \L_m(\wh A) \\
\end{split}\eeq
Although these equations are not obtained from a consistent derivation, recalling that for small $\wh A$
we have $\GG_m(\wh A) \sim \sqrt{\wh A} G_1(m)$ and $G_1(m) = 2 Q_0(m)$,
they reproduce the small cage expansion
of Ref.~\cite{PZ10} at the leading order in $\wh A$.
Note that when expressed in terms of $\wh A$ and $m$, 
the equation for the stability $\l_R=0$ is exactly the same as in $d\to\io$, Eq.~\eqref{FmLm}. Hence the result
for $\wh A_G(m)$ is independent of dimension. 

\begin{table}
\begin{tabular}{|cccc|}
\hline
$d$ & $\f_G$ & $\f_{GCP}$ & $p_G$ \\
\hline
3   &    0.683581   &   0.683657  & 26727 \\
4   &    0.486755  &    0.486874 & 16374 \\
5     &  0.330586   &   0.330718 & 12535 \\
6    &   0.218074 &    0.218203 &  10119 \\
7    &   0.140074  &   0.140189 &  8469 \\
8    &   0.0876190    &  0.0877137 &  7407 \\
9   &    0.0534490  &    0.0535198 & 6805  \\
10   &   0.0318889   &   0.0319377 & 6550   \\
11   &   0.0186760   &   0.0187075 & 6548   \\
12    &  0.0107756 &     0.0107949 & 6724   \\
13   &   0.00614419  &   0.00615559  &      7019 \\
\hline
\end{tabular}
\caption{
Values of $\f_{GCP}$~\cite{PZ10} and of $\f_G$ and $p_G$ for several dimensions.
Note that these values correspond to the {\it equilibrium} Gardner transition. For out-of-equilibrium 
states (which are the one produced in all experiments and numerical simulations),
we expect that the Gardner instability will happen at lower, possibly much lower, pressures and densities.
}
\label{tableG}
\end{table}

We will check {\it a posteriori} that even in $d=3$ the Gardner transition happens at very large pressure, hence
$m$ and $\wh A$ are small. So we can use the asymptotic expansions to obtain quantitative estimates.
The procedure is the following:
\begin{itemize}
\item Recall that at small $m$ we have $\sqrt{\wh A_G} \approx 0.8 m$.
\item Now we obtain $\f_G(m)$ (or better $m_G(\f)$) by solving
\beq
\frac{d}{2^d \f y_{liq}(\f)} = \FF_m(\wh A) \sim \frac{G_1(m)}{1-m} \frac{ \sqrt{\wh A_G} }2 \approx \sqrt{\frac\pi{m}} \, 0.4 m \approx 0.71 \sqrt{m}
\hskip10pt
\Rightarrow
\hskip10pt
m_G \approx \left(\frac{d}{0.71 \times 2^d \f y_{liq}(\f)} \right)^2 \ .
\eeq
\item
We recall from the analysis of Ref.~\cite{PZ10} that
\beq
m^* \sim \mu \, (\f_{GCP} - \f) \ ,
\hskip20pt
\mu = \frac1d \left[ 2^{d-1} y_{liq}(\f) - d \frac{y'_{liq}(\f)}{y_{liq}(\f)} + \frac{1-d}{\f} \right]_{\f = \f_{GCP}} \ .
\eeq
\item 
The Gardner transition happens when the two lines cross, hence $\f_G$ is the solution of $m_G(\f) = m^*(\f)$ which can be easily
found numerically once an equation of state for the liquid has been chosen. Here we use the Carnahan-Starling equation already
used in Ref.~\cite{PZ10}.
\item
Finally we use the result~\cite{PZ10}
\beq
p(\f_G) \sim \frac{d \, \f_{GCP}}{\f_{GCP} - \f} 
\eeq
to estimate the Gardner pressure $p_G = p(\f_G)$.
\end{itemize}

The numerical values of the Gardner pressure are reported in Tab.~\ref{tableG}. Note that the non-monotonicity of $p_G$ at
low dimension could be an artifact of the approximations used above. Note also that $p_G$ is always much larger than the pressure
at the glass transition (reported in Ref.~\cite{PZ10,CIPZ11}). 
Still, the reader should keep in mind that the value of $p_G$ corresponds
to the Gardner instability of the equilibrium (``ideal'') glass. According to the phase diagram of Fig.~\ref{pspi}, we expect
that the Gardner instability for the metastable states that are reached out of equilibrium will happen at lower pressures.
Unfortunately, quantifying this effect requires the use of ``state following'' techniques~\cite{KZ10} and goes far beyond the
scope of this article.

The limit $d\to\io$ is recovered as follows. Recall that $\wh \f_{GCP} \sim \log d$~\cite{PZ10}.
Moreover, $y_{liq} \to 1$. Hence $\mu \sim  2^{d-1}/d$ and the equation for $\wh \f_G$ becomes
\beq
\frac12 (\wh\f_{GCP} - \wh\f_G) = \left(\frac{1}{0.71 \times \wh\f_G } \right)^2 \sim  \left(\frac{1}{0.71 \times \wh\f_{GCP}}  \right)^2 \sim  \left(\frac{1}{0.71 \times \log d } \right)^2  \ ,
\eeq
which shows that the distance between $\wh \f_G$ and $\wh \f_{GCP}$ shrinks as $(\log d)^{-2}$ and the Gardner pressure diverges as
$p_G \sim d (\log d)^3$, as was sketched in Fig.~\ref{pspi}.

Finally, at this level of approximation, it can be easily shown that $\l_{\rm MCT}$ does not depend on dimension.
The small dependence of $\l_{\rm MCT}$ reported in Ref.~\cite{CIPZ12} should be explained by corrections to this approximation, that
have been neglected here.

\section{Conclusions}
\label{sec:V}

In this paper we were able to investigate the possibility of a Gardner transition for hard spheres in large spatial dimensions.
Such a study was never been done before for particle systems, and was possible thanks to the  expression of the entropy in terms of the overlap
matrix obtained in the first paper of this series, and extended here to obtain Eqs.~\eqref{eq:gauss_r} and \eqref{eq:gauss_a}: 
because this expression has been shown to be exact, any discrepancy or instability
is only attributable to the instability of the 1RSB ansatz.

The 1RSB solution
is unstable in the equilibrium glass phase for (reduced) pressure higher than the {\it Garner pressure} $p_G$~\cite{Ga85},
and for metastable states in  large region of the phase diagram, just as in $p$-spin glasses~\cite{MR03,MR04}. We provided an 
estimate of the Gardner pressure in finite dimensions, finding that it is quite high; moreover we showed that $p_G$ diverges
(slowly) with increasing dimension. These pressures are directly accessible to numerical simulations, hence we expect that
this transition should be quite easy to detect numerically. Estimating analytically the transition pressure for metastable glasses
would be very useful to guide numerical simulations: this
is however hard, as it requires a {\it state following} computation~\cite{KZ10}. Although this is possible in principle,
we leave it for future work. We expect that in any case the instability will happen at lower pressure for metastable glasses than
for the ideal glass.

The physical consequences of this instability are very intriguing but for the moment not all its implications have been worked out. 
In fact, even for the simplest $p$-spin glasses, the impact of the Gardner instability on the out-of-equilibrium dynamics
is not completely understood from a technical point of view~\cite{MR04}. The structure of the metastable states of the $p$-spin 
glass model and its impact on the out-of-equilibrium dynamics are being 
actively investigated~\cite{Ga85,MR03,MR04,CLR05,TTK07,KZ10,ZK10,Rizzo} and making progress on this simpler model will be crucial
for understanding the technically more involved hard sphere case. We expect (hope) that the scenario we proposed in Sec.~\ref{sec:general}
will be confirmed by these studies.

Let us recall here some speculations on the possible impact of the Gardner instability on the physics of jamming that we discussed in
this paper, leaving a more detailed investigation for future works.
\begin{itemize}

\item
It is reasonable to expect that at the Gardner transition, the 1RSB
solution will transform continuously into a full RSB solution, although we have not yet constructed this explicitly. 
Such a solution describes a situation where glassy states are
arranged in a complex and correlated pattern~\cite{MPV87}. More importantly, they are {\it marginally stable} ~\cite{BM79,MPV87}. This means that
the spectrum of vibrations around a glassy state displays many soft modes. Hence, it is likely that a full RSB description
of the problem will allow one to obtain information on the soft modes that are observed at the jamming transition~\cite{Wyart,LNSW10,He10}, especially those of frequency {\em below the gap} $\omega^*$.
Some steps in this direction have been already performed in Ref.~\cite{Wy12}, where the analogy with the full RSB physics of the
Sherrington-Kirkpatrick model was noticed.

\item There should be several signatures of the Gardner transition.
Suppose that the hard sphere system is prepared in a glass state in the region where the 1RSB solution is stable,
and that pressure in slowly increased approaching the instability. As mentioned in Sec.~\ref{sec:general},   
the spin glass susceptibility~\cite{MPV87}  (which in this context is a four-point static susceptibility)
diverges on approaching the instability. Moreover, even if the system was already equilibrated in
the initial glass state, aging effects should appear below the instability when the state breaks down into many correlated
sub-states.

\item
The aging curves, and in particular the fluctuation-dissipation plots, should give a good indication of the transition:
at pressures above the Gardner pressure these plots
should crossover from two straight lines to two straight lines joined by a curved segment (although detecting the curved
part might be numerically challenging).

\item There is probably a relation between the Gardner transition and the ``dynamic criticality'' defined in Ref.~\cite{IBB12}. As noticed
there, all the anomalous scalings at the jamming transition are related to the scaling of the cage radius with pressure, 
$A\sim p^{-3/2}$. Hopefully, this scaling, which is not found in the 1RSB solution, could be a property of the full RSB phase.
In fact, it is well known that in the Sherrington-Kirkpatrick the presence of a full RSB phase changes the scaling of the overlap at 
low temperatures.

\item Brito and Wyart~\cite{BW09b} have demonstrated that close to jamming, the dynamics is characterized by sudden ``cracks''
at which the system leaves abruptly a locally stable structure to find a new one. At these cracks, the displacements of the particles
are strongly correlated with the lowest frequency eigenvectors of the stability matrix of the structure that the system is leaving.
As mentioned in the introduction, this is not what is expected in a 1RSB phase, where states are locally stable and one has to cross a barrier to jump from one state
to the other -- hence the vibrations at the bottom of the well should give no information on the shape of the barrier. 
However, in a full RSB phase the dynamics is much different and similar to the one found in Ref.~\cite{BW09b}. It would be very nice
to check whether the results of Brito and Wyart really fit into a full RSB picture, for example by calculating the spatial
distribution of the displacements between two nearby states.

\item Finally, the response of full RSB magnetic systems to an external perturbation is very complex, being characterized by avalanches
and intermittency, see e.g. Ref.~\cite{LMW12}. This is due to the existence of (relatively low) barriers separating nearby states -- 
all this within a large basin.
By analogy, we would expect the response of a hard sphere system to a mechanical perturbation
in the full RSB phase to be similarly complex. Hence, the rheological properties in this phase could be very interesting and could explain
some of the anomalous behavior found around the jamming transition.
Analytical computations might be possible following the strategy introduced in Ref.~\cite{YM10,Yo12,Yo12b}.

\end{itemize}

From the technical point of view, the next step to make progress is to investigate  the $K$RSB solutions, with $K=2,3,4,\cdots$, eventually
with $K\to\io$ that corresponds to full RSB. Following Gardner's example in the $p$-spin model, this can be done just below the Gardner transition.
 This investigation is in progress and will provide some answers to the above questions. In parallel,
numerical simulations should be performed to detect the 1RSB instability. Also, an exact solution of the dynamics, along the lines of Ref.~\cite{MK11},
could provide very useful complementary informations.

To conclude, let us mention that in this paper, from the expansion of the cubic terms around the 1RSB solution (see Section~\ref{app:MCT}),
we obtained an estimate of the mean-field dynamical critical exponents at the dynamical transition (the so-called Mode-Coupling Theory exponent 
parameter $\l_{\rm MCT}$)~\cite{Go09}. We found that for $d\to\io$, $\l_{\rm MCT} = 0.70698$, which is consistent with numerical simulations~\cite{CIPZ12}.
This is important because a previously attempted calculation from the replicated HNC equations~\cite{FJPUZ13} gives results that are 
quantitatively bad.
The fact that in $d\to\io$ we can obtain a good result implies that the negative result obtained in Ref.~\cite{FJPUZ13} has to be attributed to the poor quantitative
performances of the replicated HNC approximation, which indeed were already known~\cite{PZ10}. 
Unfortunately, also the approach presented here gives poor quantitative results for the dynamical glass transition in low dimensions~\cite{PZ10}.
Obtaining an accurate theory of the dynamical glass transition in low dimension by improving the replicated HNC is therefore very important, see~\cite{JZ12} 
for a preliminary step in this direction.

\acknowledgments
FZ wishes to warmly thank F.~Santambrogio for very useful discussions and for hospitality during part of this work,
and C.~Barbieri for a crucial help with Mathematica.
We thank P.~Charbonneau and S.~Franz for many important discussions and for collaborating in previous projects to obtain
analytical and numerical results for hard spheres which strongly motivated this study.
We also thank F.~Krzakala, F.~Ricci-Tersenghi, T.~Rizzo and L.~Zdeborov\'a for very useful discussions.
Financial support was provided by
the European Research Council through ERC grant agreement
no. 247328.

\appendix

\section{Computation of the Jacobian $J(\hat q)$}
\label{app:J}

The Jacobian $J(\hat q)$ is defined in the following way
\beq\label{appA1}
J(\hat q)=m^d\prod_{a=1}^m\delta\left[\sum_{b=1}^mq_{ab}\right]\int \de^d u_1\ldots\de^du_{m-1}\prod_{a\leq b}^{1,m-1}\delta\left(q_{ab}-u_a\cdot u_b\right)
\eeq
Let us introduce the following notation. We define a $d\times (m-1)$ matrix $U$ whose first column contains the components of the $d$-dimensional vector $u_1$, the second column contain the components of $u_2$ and the $m-1$ column contains the components of $u_{m-1}$.
Moreover we define the $m\times m$ matrix $\hat q$ and its reduced version that is the matrix $\hat q^{(m,m)}$ which is obtained from $\hat q$ by deleting the last column and the last row. Let us consider the integral in the expression~\eqref{appA1}:
\beq
\int \de^d u_1\ldots\de^du_{m-1}\prod_{a\leq b}^{1,m-1}\delta\left(q_{ab}-u_a\cdot u_b\right)=\int \de U \delta\left[\hat q^{(m,m)}-U^TU\right] \ ,
\eeq
where the last integral is with a flat measure over the entries of the matrix $U$.
To perform this computation we start from a simple case. Let us consider the case where the matrix $\hat q^{(m,m)}$ has a diagonal structure
\beq
\hat q^{(m,m)}\equiv \hat q_D \  , \hskip20pt (\hat q_D)_{ab} =q_{aa}\delta_{ab} \ .
\eeq
This means that
\beq
\int \de U \delta\left[\hat q^{(m,m)}-U^TU\right]=\int \de^d u_1\ldots\de^du_{m-1}\prod_{a=1}^{m-1}\delta\left(q_{aa}-|u_a|^2\right)\prod_{a<b}^{(1,m-1}\delta(u_a\cdot u_b) \ .
\eeq
The second Dirac delta function says that the vectors $u$ have to be all orthogonal one to the other. The first one tells us about the length of these vectors. By going to polar coordinates we have that the previous expression is given by
\beq
\int \de^d \hat u_1\ldots\de^d\hat u_{m-1}\int_0^\infty \de u_1 u_1^{d-1}\ldots \int_0^\infty \de u_{m-1} u_{m-1}^{d-1}\prod_{a=1}^{m-1}\delta\left(q_{aa}-u_a^2\right)\prod_{a<b}^{1,m-1}\delta(u_au_b \hat u_a\cdot \hat u_b)
\eeq
where we have denoted with $\de^d\hat u$ the angular integration in $d$ dimensions and with $\hat u_a$ the unit vector parallel to $u_a$. Rewriting the Dirac deltas in different ways the previous expression can be written in the following way
\beq\begin{split}
2^{1-m}\int \de^d \hat u_1\ldots& \de^d\hat u_{m-1} \int_0^\infty \de u_1 u_1^{d-1}\ldots \int_0^\infty \de u_{m-1} u_{m-1}^{d-1}\left(\prod_{a=1}^{m-1}\frac{1}{\sqrt{q_{aa}}}\delta\left(\sqrt{q_{aa}}-u_a\right)\right) \times \\
&\times \left(\prod_{a<b}^{1,m-1}\frac{1}{\sqrt{q_{aa}q_{bb}}}\right)\prod_{a<b}^{1,m-1}\delta(\hat u_a\cdot \hat u_b) 
=C_{m,d}\prod_{a=1}^{m-1}q_{aa}^{(d-m)/2}=C_{m,d}\left[\det \hat q^{(m,m)}\right]^{(d-m)/2} \ ,
\end{split}\eeq
where we have
\beq
C_{m,d}=2^{1-m}\int \de^d \hat u_1\ldots\de^d\hat u_{m-1}\prod_{a<b}^{1,m-1}\delta(\hat u_a\cdot \hat u_b) \ .
\eeq
The integral that appears in the expression for $C_{m,d}$ is very simple. In fact the Dirac deltas tell us that the unit vectors $\hat u$ must be all orthogonal. So we have to compute the phase space accessible to them. This can be done iteratively. Suppose that we have only one unit free vector. It has a phase space available given by the solid angle in $d$ dimensions which is $\Omega_d$. Then we add a second unit vector orthogonal to the first one. Clearly it has a phase space available that is the solid angle in the space orthogonal to the first vector that is $\Omega_{d-1}$. Going on by iteration we have
\beq
C_{m,d}=2^{1-m}\Omega_d\Omega_{d-1}\ldots \Omega_{d-m+2}.
\eeq
Now we want to generalize this to a matrix $\hat q^{(m,m)}$ that is not diagonal.
Because the matrix $\hat q^{(m,m)}$ is symmetric we can diagonalize it and we can write
\beq
\hat q^{(m,m)}=\Lambda^{-1}\hat q_D\Lambda \ , \ \ \ \ \ \ \ 
\det \Lambda=1 \ , \ \ \ \ \ \ 
\det \hat q_D=\det \hat q^{(m,m)} \ ,
\ \ \ \ \Lambda^T=\Lambda^{-1} \ .
\eeq
It follows that
\beq
\int \de U \delta\left[\hat q^{(m,m)}-U^TU\right]=\int \de U \delta\left[\Lambda^{-1}\hat q_D\Lambda-U^TU\right]=\int \de U \delta\left[\hat q_D-\Lambda U^TU\Lambda^{-1}\right] \ .
\eeq
If we change integration variables
\beq
U'=U\Lambda^{-1} \ , 
\eeq
that has a Jacobian which is unitary due to the orthogonality property of the matrix $\Lambda$, then the previous expression becomes
\beq
\int \de U \delta\left[\hat q_D-\Lambda U^TU\Lambda^{-1}\right]=\int \de U \delta\left[\hat q_D-U^TU\right] \ ,
\eeq
that is the same that we discussed above. It follows that
\beq
J(\hat q)=m^dC_{m,d}\prod_{a=1}^m\delta\left[\sum_{b=1}^mq_{ab}\right]\left[\det \hat q^{(m,m)}\right]^{(d-m)/2} \ ,
\eeq
which is the result that was used in Ref.~\cite{KPZ12}.

\clearpage

\bibliographystyle{mioaps}
\bibliography{HS} 



\end{document}